\documentclass[floatfix,aps,pre,showpacs]{revtex4}
\usepackage{graphicx}
\usepackage{amsmath}
\usepackage{amssymb}
\usepackage{mathrsfs}

\begin{document}

\title{Communities of solutions in single solution clusters of a random $K$-Satisfiability formula}

\author{Haijun Zhou$^{1,2}$ and Hui Ma$^1$}

\affiliation{$^1$Key Laboratory of Frontiers in Theoretical Physics and
		$^2$Kavli Institute for Theoretical Physics China, 
		Institute of Theoretical Physics, Chinese Academy of Sciences,
		Beijing 100190,	China}

\date{\today}

\begin{abstract}
The solution space of a $K$-satisfiability ($K$-SAT) formula is a collection
of solution clusters, each of which contains all the solutions
that are mutually reachable through a sequence of single-spin flips.
Knowledge of the statistical property of solution clusters
is valuable for a complete understanding of the
solution space structure and the computational complexity
of the random $K$-SAT problem.
This paper explores single solution clusters of random $3$- and $4$-SAT
formulas through unbiased and biased random walk processes
and the replica-symmetric cavity method of statistical physics. We find
that the giant connected component of the
solution space has already formed many
different communities when the constraint density
of the formula is still lower than the
solution space clustering transition point.
Solutions of the same community are more similar with each other and
more densely connected with each other
than with the other solutions. The entropy density of
a solution community is calculated using belief propagation
and is found to be different for different communities of the
same cluster. When the constraint density is beyond the clustering
transition point, the same behavior is observed for
the solution clusters reached by several
stochastic search algorithms. Taking together, the results of this work 
suggests a refined picture on the evolution of the solution space
structure of the random $K$-SAT problem; they may also be helpful for
designing new heuristic algorithms.
\end{abstract}


\maketitle

\section{Introduction}
\label{sec:introduction}

As the $`$Ising model' of intrinsically hard combinatorial
satisfaction problems, the random $K$-satisfiability ($K$-SAT) problem
was extensively studied in the last twenty years.
Recent major progresses include mean-field predictions
and rigorous bounds on the satisfiability threshold
\cite{Mezard-etal-2002,Biroli-Monasson-Weigt-2000,Achlioptas-Naor-Peres-2005},
mean-field predictions on various structural transitions in
the solution space of a random $K$-SAT formula
\cite{Krzakala-etal-PNAS-2007}, and new efficient stochastic algorithms
\cite{Mezard-etal-2002,Selman-Kautz-Cohen-1996,Alava-etal-2008}.
Statistical physics theory
\cite{Mezard-etal-2002,Mezard-etal-2005,Krzakala-etal-PNAS-2007} predicted that the
solution space of a satisfiable
random $K$-SAT formula ($K\geq 3$) divides into exponentially many
Gibbs states as the constraint density is beyond a clustering (dynamic)
transition point. For $K \geq 8$ it was proved \cite{Mezard-etal-2005-a} that
the solution space Gibbs states are extensively separated from each other, but
whether the same picture holds for $3 \leq K < 8$ is still an open question.
Recent empirical studies revealed that for random $K$-SAT formulas with $K<8$
the clustering transition has no fundamental restriction  on the
performances of some stochastic search algorithms such as {\tt WALKSAT} and
{\tt ChainSAT} \cite{Seitz-Alava-Orponen-2005,Alava-etal-2008}.
For example,  the {\tt ChainSAT} process \cite{Alava-etal-2008}
is able to find solutions for a random $4$-SAT formula
with constraint density well beyond the clustering transition value,
although during the search process the number of unsatisfied constraints
of the formula never increases. The most efficient
stochastic algorithm for large random $K$-SAT formulas is
survey propagation \cite{Mezard-etal-2002} which, for the random $3$-SAT
problem, is able to
find solutions at constraint densities extremely chose to the satisfiability
threshold. To understand the high efficiency of these
and other stochastic search algorithms,
it is desirable to have  more detailed knowledge
on the energy landscape and the solution space
structure of the random $K$-SAT problem 
(see, e.g., Refs.~\cite{Krzakala-Kurchan-2007,Ardelius-Zdeborova-2008}
for some very recent efforts). Such knowledge will also be very helpful
for designing new stochastic search algorithms.

A random $K$-SAT formula contains $N$ variables and $M = \alpha N$ clauses,
$\alpha$ ($\equiv M /N$) being the constraint density.
Each variable has a spin $\sigma =\pm 1$, and
each clause prohibits $K$ randomly chosen variables
from taking a randomly specified spin configuration of the $2^K$ possible
ones.  The configurations $\vec{\sigma} \equiv 
\{ \sigma_1, \ldots, \sigma_N\}$ that  satisfy a  formula $F$
forms a solution space. The Hamming distance of two solutions is defined as
\begin{equation}
\label{eq:distance}
	d( \vec{\sigma}^1, \vec{\sigma}^2 )
	 = \sum\limits_{j=1}^{N} \delta(\sigma_j^1, -\sigma_j^2) \ ,
\end{equation}
where $\delta(x, y)=1$ if $x=y$ and $0$ otherwise.
Two solutions $\vec{\sigma}^1$ and
$\vec{\sigma}^2$ are regarded as nearest neighbors if
they differ on just one variable, i.e., $d(\vec{\sigma}^1, \vec{\sigma}^2)=1$.
The organization of the solution space can be studied graphically by
representing each solution as a vertex and connecting every  pair of
unit-distance solutions
by an edge. Then the solution space can be regarded as a collection of
solution clusters, each of which is a connected component of the solution
space in its graphical representation.
How many solution clusters 
does this astronomically huge graph contain? What is the size distribution of
these clusters? 
What are the distributions of the minimal, the mean, and the maximal
distances between two
clusters? How are the solutions in each cluster organized?
These questions are fundamental to a complete understanding of the random
$K$-SAT problem, but they are very challenging and 
so far only few rigorous mathematical answers are achieved
\cite{Achlioptas-Naor-Peres-2005,Mezard-etal-2005-a}.
Mean-field statistical physics theory
\cite{Mezard-etal-2005,Krzakala-etal-PNAS-2007}
is able to give a prediction on the number of solution Gibbs states of
a given size, but whether there is a strict one-ton-one
correspondence between solution Gibbs states, which are defined according
to statistical correlations of the solution space
\cite{Montanari-Semerjian-2006,Montanari-Semerjian-2006b}, and solution clusters
is not yet completely clear. 

Following our previous work Ref.~\cite{Li-Ma-Zhou-2009} in this paper
we focus on one of the structural aspects of the solution
space, namely the organization of
a single connected component (a solution cluster).
The internal structure of a solution cluster is explored by 
unbiased and biased random walk processes.
We examine mainly solution clusters reached by a
very slow
belief propagation decimation algorithm, but it appears that
the qualitative results
are the same for solution clusters reached by various other algorithms.
We can verify that the studied solution
clusters correspond to the single (statistically relevant) Gibbs state
of the given formulas if the constraint density $\alpha$ is lower than
$\alpha_d$, the clustering transition point where 
exponentially many Gibbs states emerge \cite{Krzakala-etal-PNAS-2007}.
We find that the solutions in such a 
giant cluster already aggregate into many
different communities when $\alpha$
is still much lower than $\alpha_d$.
In a solution cluster, solutions of the same community are
more densely connected with each other than with the other solutions,
and the mean Hamming distance of
solutions belonging to the same community is shorter than the mean
solution-solution Hamming distance of the whole cluster.
The entropy density of
a solution community is calculated by the replica-symmetric
cavity method of statistical
physics and is found to be different for different communities of the
same cluster. When the constraint density exceeds
$\alpha_d$, we have the same observation that
non-trivial community structures are present in the single solution clusters
reached by several stochastic search algorithms.
These numerical results are interpreted in terms of
the following proposed evolution picture
of the solution space of a random $K$-SAT formula: (1)
As the number of constraints of the formula increases and $\alpha$ becomes
close to $\alpha_d$ from below, many
relatively densely connected solution communities
 emerge
in the solution spaces and
these communities are linked to each other by
various inter-community edges; 
(2) the intra- and inter-community connection patterns both evolve
with $\alpha$, and finally the single giant component of the solution space
breaks into many clusters of various sizes (probably at $\alpha=\alpha_d$),
each of which 
contains a set of communities; (3) as $\alpha$ further increases,
the intra- and inter-community
connection patterns in each solution cluster keep evolving, 
leading to the breaking of a solution cluster into sub-clusters.

The following section describes the numerical methods used in this paper.
The simulation results on random $3$-SAT and $4$-SAT formulas
are reported in Sec.~\ref{sec:3sat} and Sec.~\ref{sec:4sat}, respectively.
We conclude this work in Sec.~\ref{sec:conclusion}.

\section{Methods}
\label{sec:methods}

\subsection{The random walk processes and the data clustering method}

\begin{figure}
	\includegraphics[width=0.5\linewidth]{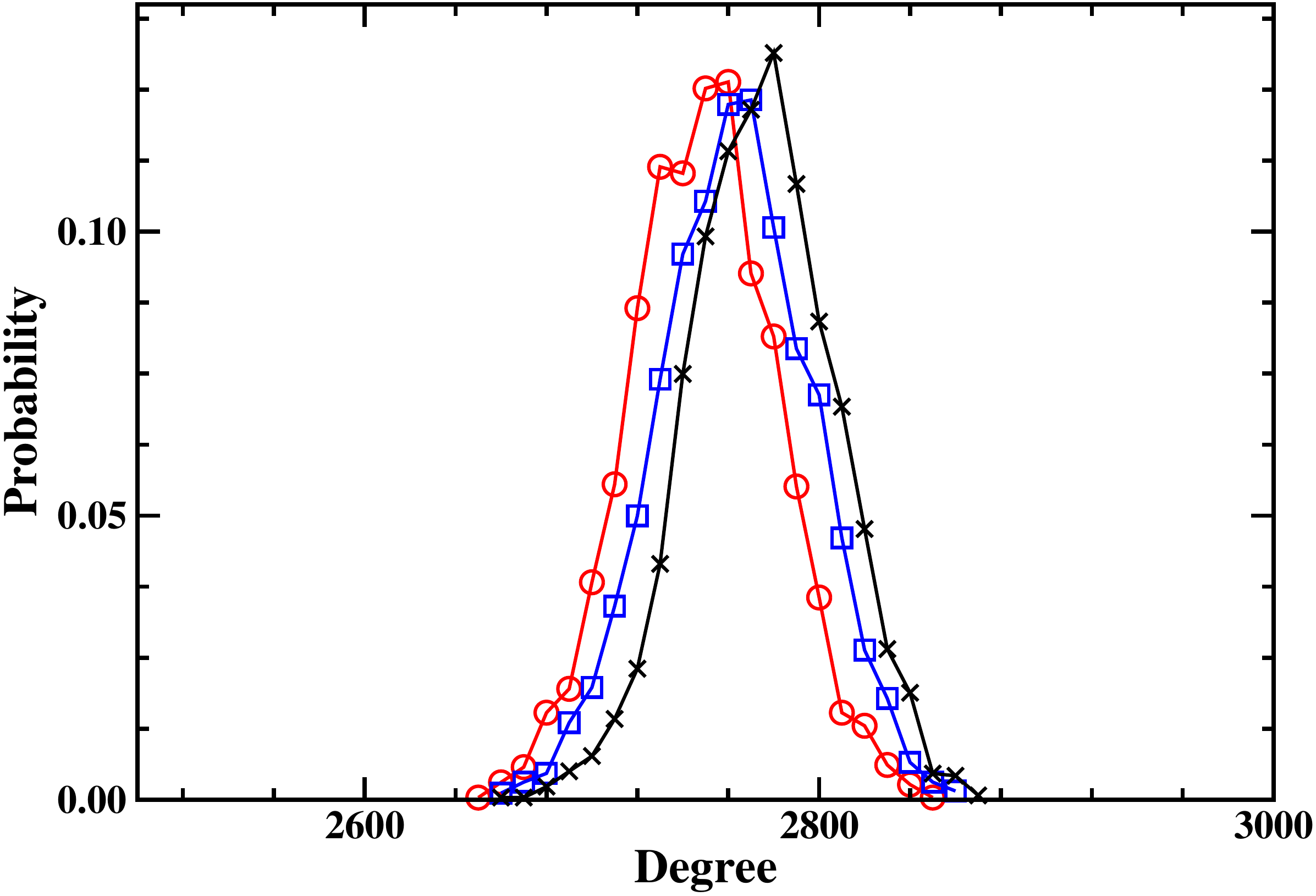}
	\caption{\label{fig:degreeprofile}
	(Color online) The degree distribution of solutions from
	a solution cluster. The three curves correspond to
	three random $3$-SAT formulas of $N=20,000$ variables
	and constraint density $\alpha=3.925$. To get a
	degree distribution, $2,500$ solutions
	are {\em uniformly sampled} from a solution cluster
	by a Markov chain process.
	Suppose at time $t$ the solution $\vec{\sigma}=\{\sigma_1, \ldots,
	\sigma_i, \ldots, \sigma_N\}$ is being visited.
	A variable $i$ is chosen with probability
	$1/N$ from the whole set of variables.
 	If this variable can be flipped without violating any constraint
	of the formula,	it is flipped and the solution is updated to
	$\vec{\sigma}^\prime = 
	\{ \sigma_1, \ldots, -\sigma_i, \ldots, \sigma_N\}$
	at time $t^\prime = t+ \delta$, otherwise
	the old solution $\vec{\sigma}$ is kept at
	time $t^\prime$. We set $\delta  = 1/N$ and
	sample solutions at an equal time interval of $80,000$.
	}
\end{figure}

\begin{figure}
	\includegraphics[width=0.9\textwidth]{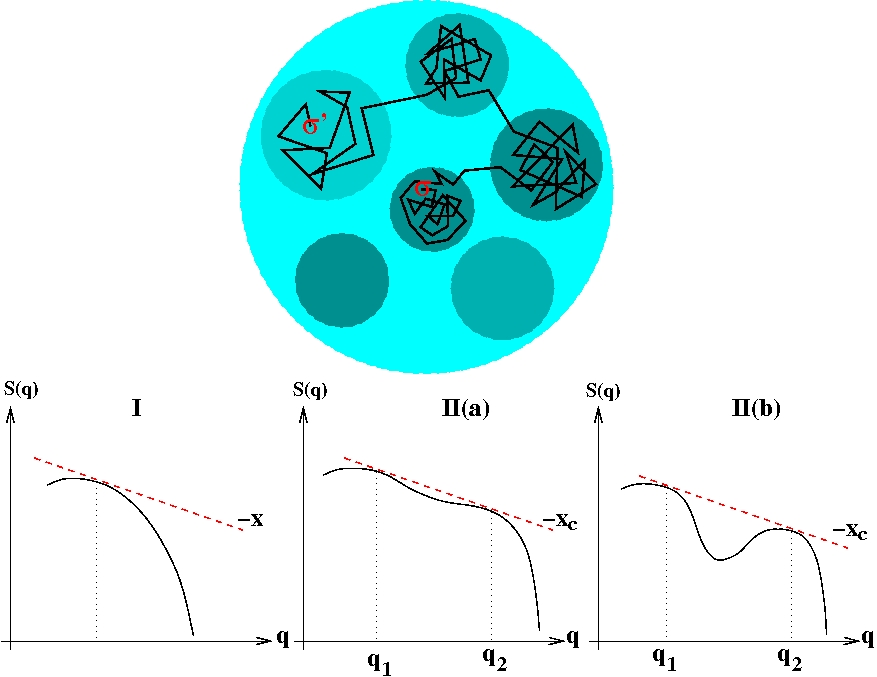}
	\caption{\label{fig:communityschematic}
	(Color online) (upper panel) Schematic view of
	solution communities in a single solution cluster.
	The mean edge density in the whole cluster (the
	largest circle) is less than the edge
	densities of individual communities  (small circles).
	A path of single-spin flips linking solutions
	$\vec{\sigma}$ and $\vec{\sigma}^\prime$ of two different communities
	is shown by the black coiled trajectory.
	(lower panel) Entropy density $s(q)$ as a function of
	 the overlap $q$ with
	a given reference solution. If $s(q)$ is a concave function
	(case I), a rectilinear line with slope $-x$ can only
	be tangent to $s(q)$ at one point;
	if $s(q)$ is  not concave, then
	a rectilinear line with certain slop $-x_c$ may be tangent to
	$s(q)$ at two points $q_1$ and $q_2$. In the 
	interval of $q_1 \leq q \leq q_2$, $s(q)$ may be monotonic [case II(a)]
	or be non-monotonic [case II(b)].
	}
\end{figure}

A solution cluster contains a huge number ${\cal{N}} \sim \exp(N s)$
of solutions, with $s$ being the
entropy density. A solution $\vec{\sigma}$ in this cluster
is connected to $k_{\vec{\sigma}}$ other solutions,  
$1 \leq k_{\vec{\sigma}} \leq N$. Empirically we
found that the degrees $k_{\vec{\sigma}}$ of the solutions in a cluster 
are narrowly distributed with a mean much less than $N$
(see Fig.~\ref{fig:degreeprofile} for an example).
Therefore the solutions of a
cluster can be regarded as almost equally important in terms of connectivity.
However, the connection pattern of the solution cluster can be
highly heterogeneous.
Solutions of a cluster may form different communities such that
the edge density of a community  is much larger that of the whole
cluster (Fig.~\ref{fig:communityschematic} (upper panel)
gives a schematic picture, where darker circles indicate solution
communities with higher edge densities).
The communities may even further organize into
super-communities to form a hierarchical structure.
If a random walker is following the edges of such a community-rich
solution cluster, it will be trapped in different communities
most of the time and only will spend a very small fraction of its time
traveling between different communities. If solutions are sampled by the
random walker at equal time interval $\Delta t$, the sampled solutions
contains useful information about the community structure of the
solution cluster at a resolution level that depends on
$\Delta t$.


Two slightly different random walk processes are used in this paper
to explore the structure of single solution clusters.
The first one is {\tt SPINFLIP} of Ref.~\cite{Li-Ma-Zhou-2009},
which prefers to flip newly discovered unfrozen variables.
Starting from an initial solution denoted as $\vec{\sigma}^*$ at time $t=0$,
the {\tt SPINFLIP} process explores a solution cluster by
jumping between nearest-neighboring solutions.
The set $U$ of discovered unfrozen (flippable) variables is initially empty.
Suppose the walker resides on $\vec{\sigma}(t)$ at time $t$.
The set of flippable variables in this solution is divided into two sub-sets:
set $A(t)$ contains all the variables that have already
been flipped at least once, set $B(t)$ contains the remaining flippable
variables.
In the time interval $\delta = 1/N$ the spin of a randomly chosen variable
in set $B(t)$ (if $B(t) \neq \varnothing$) or set $A(t)$ (if otherwise)
is flipped.
At time $t^\prime=t+\delta$ the walker is then in a nearest-neighbor
of $\vec{\sigma}(t)$,
and the updated set of unfrozen variables is $U(t^\prime)=U(t)\cup B(t)$.
A unit time of {\tt SPINFLIP} corresponds to $N$ flips.
As newly discovered unfrozen variables are flipped by
{\tt SPINFLIP} with priority, the random walker probably can escape from the
local region of the initial solution $\vec{\sigma}^*$  quicker than
an unbiased random walker.
However we have checked that this slight bias is not
at all significant to the simulation results. There are two reasons: first
the random walk process occurs in a high-dimensional space, and second, after
a brief transient time the set $B(t)$ of newly discovered unfrozen variables
becomes empty most of the time.

We also use the unbiased random walk process in some of the simulations.
The unbiased random walk differs from {\tt SPINFLIP} in that at each
elementary solution update, a variable is uniformly randomly
chosen from the set of flippable variables and flipped. As we just mentioned,
{\tt SPINFLIP} converges to the unbiased random walk as the
simulation time $t$ becomes large enough (e.g., $t \approx 10^6$).

A number of solutions are
sampled with equal time interval 
$\Delta t$ during the random walk process for clustering
analysis. The overlap $q$ between any two
sampled solutions $\vec{\sigma}^1$ and $\vec{\sigma}^2$ is defined by
\begin{equation}
	\label{eq:overlap}
	q(\vec{\sigma}^1, \vec{\sigma}^2)
	\equiv 1- \frac{2 d( \vec{\sigma}^1, \vec{\sigma}^2)}{N} \ .
\end{equation}
We can obtain an overlap histogram from the sampled solutions.
A hierarchical minimum-variance 
clustering analysis \cite{Jain-Dubes-1988}
is performed on these sampled solutions (the same method  was used
by Hartmann and co-workers to study the ground state-spaces
of some optimization problems 
\cite{Barthel-Hartmann-2004}). Initially each solution is
regarded as a group, and the
distance between two groups is just the Hamming distance.
At each step of the clustering, two groups $C_a$ and $C_b$
that have the smallest distance are merged into a single group $C_c$.
The distance between $C_c$ and another group $C_d$ is calculated by
\begin{equation}
	\label{eq:clusterdistance}
	d(C_c, C_d) = \frac{(|C_a| + |C_d|) d(C_a, C_d)
 	+ (|C_b|+|C_d|) d(C_b, C_d) - |C_d| d(C_a, C_b)}{|C_c|+ |C_d|} \ ,
\end{equation}
where $|C|$ denotes the number of solutions in group $C$.
A dendrogram of groups is obtained from this clustering
analysis, and the matrix of Hamming distances
of the sampled solutions is drawn with the solutions being ordered
according to this dendrogram \cite{Barthel-Hartmann-2004}.

We should emphasize that, by the above-mentioned random walk processes,
solutions of a cluster are sampled with probability proportional to
its connectivity rather than with equal probability.  We can also
sample solutions uniformly random by a slight change of the random walk
process as explained in the caption of Fig.~\ref{fig:degreeprofile}.
We have checked that the results of this paper are not qualitatively
changed by this different sampling method. This may not be surprising:
for one hand, the degrees of different solutions of the same
cluster are very close to each other, and for the other hand, if there
is many communities in a solution cluster, their trapping effects will
be felt by different random walk processes.

\subsection{Entropy calculation using the replica-symmetric cavity method}

For a solution community, some of the important statistical quantities
are the entropy density, the mean overlap between two solutions of the
community, and the mean overlap between a solution of the
community and a solution outside of the community. The entropy density
$s$ is  defined by
\begin{equation}
	\label{eq:entopydensity}
	s \equiv \frac{\ln (\mathcal{N})}{N} \ ,
\end{equation}
where $\mathcal{N}$ is the number of solutions in the community.
Following Ref.~\cite{DallAsta-etal-2008} we use the
replica-symmetric cavity method of statistical physics
 \cite{Mezard-Parisi-2001} to evaluate the
values of these quantities. The replica-symmetric cavity method is
equivalent to the belief propagation (BP) method of computer science
\cite{Pearl-1988}.

Suppose $\vec{\sigma}^1$ is a sampled solution from
a solution community. With respect to this solution,
a partition function $Z(\vec{\sigma}_1, x)$ is defined as
\begin{equation}
	\label{eq:partitionfunction}
	Z(\vec{\sigma}^1, x)
	= {\sum\limits_{\vec{\sigma}}}^\prime
	\exp\Bigl[ N x \sum\limits_{j=1}^N \sigma_j^1 \sigma_j \Bigr]
	 = {\sum\limits_{\vec{\sigma}}}^\prime
	 \exp\bigl[ N x  q(\vec{\sigma}^1, \vec{\sigma})\bigr] \ ,
\end{equation}
where $\sum^\prime$ means that only the solutions of the
formula are summed. When the reweighting parameter $x=0$, all solutions
contribute equally to the partition function $Z(\vec{\sigma}^1,0)$, which
is just equal to the total number of solutions. At the other limit of
$x\gg 0$, only those solutions $\vec{\sigma}$ with
$q(\vec{\sigma}^1, \vec{\sigma}) \approx 1$ contribute
significantly to $Z(\vec{\sigma}^1,x)$. At a given value of $x$,
Eq.~(\ref{eq:partitionfunction}) can be expressed as
\begin{equation}
	\label{eq:partitionfunction-2}
	Z(\vec{\sigma}^1, x)= \sum\limits_{q} \exp\Bigl[ N \bigl( s(q) + x q\bigr) \Bigr] \ ,
\end{equation}
where $e^{N s(q)}$ is the total number of solutions whose overlap
value with $\vec{\sigma}^1$ is equal to $q$.  
$s(q)$ is referred to as the entropy
density of solutions at overlap value $q$. 
When $N$ is large, the summation of Eq.~(\ref{eq:partitionfunction-2}) is contributed
almost completely by the terms with the maximum value of the
function $f(q,x)\equiv s(q)+ x q$. At a given $x$,
the relevant overlap value $q$ to
$Z(\vec{\sigma}_1,x)$ is therefore determined by
\begin{equation}
	\label{eq:qvariation}
	{\frac{{\rm d} s(q)}{{\rm d} q}}= - x \ ,
\end{equation}
and the corresponding entropy density at this $q$ value is related to $f(q,x)$ by
a Legendre transform $s(q)= f(q,x)-x q$. The following BP iteration scheme
is used to determine the overlap and entropy density as a function of $x$.
The function $s(q)$ is then obtained from these two
data sets by eliminating $x$.

When applying the replica-symmetric cavity method to
a single random $K$-SAT formula, first one needs to define two cavity
quantities $\eta_{i\rightarrow a}$ and $u_{a\rightarrow i}$:
\begin{eqnarray}
	\eta_{i\rightarrow a} &= & \ln\Bigl[
	\frac{P_{i\rightarrow a}(+1)}{P_{i\rightarrow a}(-1)}\Bigr] \,
	\label{eq:eta-i-a} \\
	u_{a\rightarrow i} &=& \ln\Bigl[ 1-\prod\limits_{j\in \partial a
	\backslash i} P_{j\rightarrow a}(-J_a^j) \Bigr] \ .
	\label{eq:u-a-i}
\end{eqnarray}
In the above two equations, $P_{i\rightarrow a}(\sigma_i)$ is the
(cavity) probability of variable $i$ to take the spin value
$\sigma_i$ if it is not constrained by constraint $a$; 
$\partial a$ denotes the set of variables that are involved in constraint
$a$, and $\partial a \backslash i$ is identical to $\partial a$ except
that variable $i$ is missing;  $J_a^j = \pm 1$ is the satisfying spin value of
variable $i$ for constraint $a$
(i.e., $J_a^j=+1$ (respectively $-1$) if $\sigma_i=+1$
( $-1$) satisfies $a$).
The cavity quantity $u_{a\rightarrow i}$ is the
log-likelihood of constraint $a$ being satisfied by variables other than
variable $i$.

The following BP iteration equations can be written down for
$\eta_{i\rightarrow a}$ and $u_{a\rightarrow i}$
 (see, e.g.,
Refs.~\cite{Montanari-etal-2008,Zhou-2008}):

\begin{eqnarray}
\eta_{i\rightarrow a} &=& 2 x \sigma_i^1 -
\sum\limits_{b\in \partial i \backslash a}  J_b^i
u_{b\rightarrow i} \,
\label{eq:eta-i-a-iter} \\
u_{a\rightarrow i} &=&
	\ln\Bigl[1-\prod\limits_{j\in \partial a \backslash i} 
	\frac{1+J_a^j + (1-J_a^j) e^{\eta_{j\rightarrow a}}}
	{2(1+e^{\eta_{j\rightarrow a}})} \Bigr]
\ . \label{eq:u-a-i-iter}
\end{eqnarray}
In Eq.~(\ref{eq:eta-i-a-iter}), $\partial i$ denotes the set of
constraints in which $i$ is involved, $\partial i \backslash a$ is the
a subset of $\partial i$ with $a$ being removed.

After a fixed-point solution is obtained at a given value of $x$
for the set of cavity
quantities $\{\eta_{i\rightarrow a}, u_{a\rightarrow i}\}$, the
overlap $q$ is then calculated by the following equation
\begin{equation}
	q = \frac{1}{N} \sum\limits_{i=1}^{N}
\sigma_i^1 \langle \sigma_i \rangle =
\frac{1}{N} \sum\limits_{i=1}^N
	 \frac{ \sigma_i^1 ( e^{\eta_i} -1 )}{e^{\eta_i} + 1} \ ,
\end{equation}
where $\langle \sigma_i \rangle$ is the average value of $\sigma_i$
at the reweighting parameter $x$, and $\eta_i$ is equal to
\begin{equation}
	\eta_{i}= 2 x \sigma_i^1 -
	\sum\limits_{a\in \partial i}  J_a^i u_{a\rightarrow i}
\end{equation}
The entropy density is expressed as
\begin{equation}
	s = \frac{1}{N} \sum\limits_{i=1}^N \Delta S_i
	-\frac{1}{N}\sum\limits_{a=1}^{M} (K-1) \Delta S_a
	- x q \ ,
\end{equation}
where
\begin{eqnarray}
	\Delta S_i
	&=&\ln\Bigl[\exp\bigl( - x \sigma_i^1+\sum\limits_{a \in \partial i:
	J_{a}^i=1} u_{a\rightarrow i}\bigr) + \exp\bigl(x \sigma_i^1
	+ \sum\limits_{a \in \partial i: J_{a}^{i}=-1} u_{a\rightarrow i}\bigr)
	\Bigr] \ , \\
	\Delta S_a &=&
	\ln\Bigl[ 1-\prod\limits_{i\in \partial a} 
	\frac{1+J_a^i + (1-J_a^i) e^{\eta_{i\rightarrow a}}}
	{2(1+e^{\eta_{i\rightarrow a}})} \Bigr]  \ .
\end{eqnarray}

At a given value of $x$, one can also estimate the mean overlap $\bar{q}(x)$
between two solutions of the solution space by
\begin{equation}
	\bar{q}(x)=\frac{1}{N}\sum\limits_{i=1}^{N} \langle \sigma_i \rangle^2
	=\frac{1}{N}\sum\limits_{i=1}^{N} 
	\frac{(e^{\eta_i}-1)^2}{(e^{\eta_i}+1)^2} \ .
\end{equation}

As we will demonstrate in the next two sections, when the reweighting
parameter $x$ is equal to certain critical values,
the calculated entropy density $s$ and overlap $q$ may change discontinuously
with $x$. Furthermore, at certain range of the parameter $x$,
the BP iteration equations may have two fixed-points with
different $s$ values and $q$ values. Such behaviors are caused by
the non-concavity of the entropy density function $s(q)$. As shown
in Fig.~\ref{fig:communityschematic} (lower panel),
if $s(q)$ is non-concave, then at 
certain critical value $x=x_c$, Eq.~(\ref{eq:qvariation}) has two 
solutions at $q_1$ and $q_2$, with $q_1 < q_2$. When $x$ is
slightly larger than
$x_c$, we have $f(q_2, x) > f(q_1,x)$. Therefore the partition function
$Z(\vec{\sigma}^1,x)$ is dominantly contributed by solutions of overlap value
$q\approx q_2$, and the total number of
these solutions is $e^{N s(q_2)}$, while the solutions with overlap
$q\approx q_1$ form a $`$metastable' state. When $x$ is slightly smaller than
$x_c$, then $f(q_1, x) > f(q_2, x)$ and the reverse is true:
$Z(\vec{\sigma}^1, x)$ is contributed predominantly by
solutions with overlap $q\approx q_1$, and the total
number of these solutions is $e^{N s(q_1)}$, and the solutions at overlap
$q\approx q_2$ form a metastable state. At $x\approx x_c$, the two 
fixed-point solutions of the BP iteration equations
correspond to these two maximal points of $f(q, x)$.
 
The non-concavity of $s(q)$ at certain range of overlap values is
a strong indication that the solution space has non-trivial structures,
which might be the existence of many solution clusters, or the existence
of many solution communities in the solution cluster of $\vec{\sigma}^1$,
or both. The reweighting parameter $x$ in Eq.~(\ref{eq:partitionfunction})
can be regarded as an external field which biases the spin of each variable
$i$ to $\sigma_i^1$.  At the limit of $N\rightarrow \infty$, for
the non-concave cases shown in II(a) and II(b) of
Fig.~\ref{fig:communityschematic}, a real
first-order phase-transition will occur at $x=x_c$ between an 
energy-favored phase with overlap $q\approx q_2$ and an 
entropy-favored phase with overlap $q\approx q_1$.

\section{Results for random $3$-SAT formulas}
\label{sec:3sat}

\subsection{Random walk on a solution cluster reached by survey propagation}
\label{sec:3sat4p25n1m}

\begin{figure}[t]
	\includegraphics[width=0.9\linewidth]{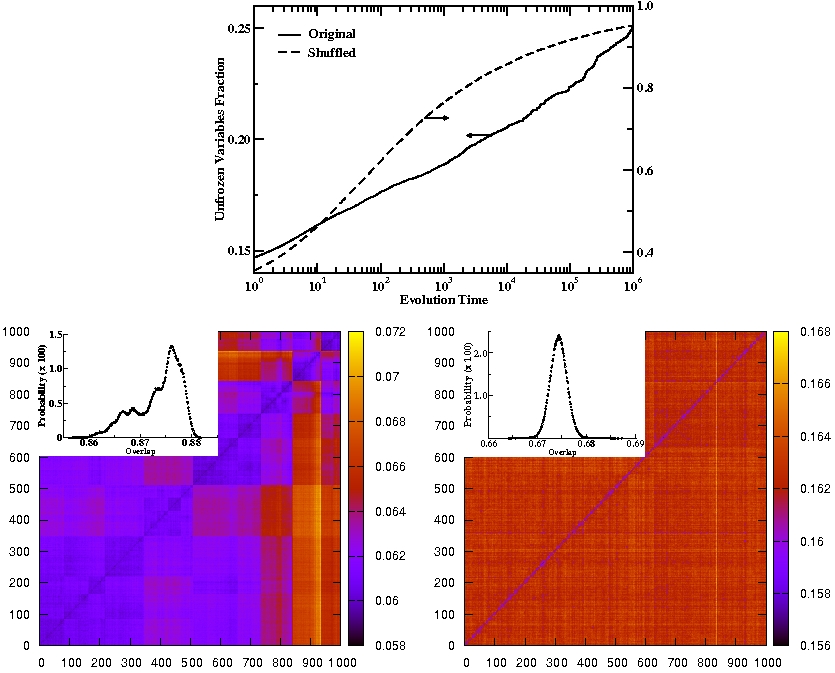}
	\caption{\label{fig:3sat4p25n1m}
	(Color online) Simulation results for a random $3$-SAT
	formula with $N=10^6$ variables and constraint density
	$\alpha=4.25$: Number of discovered unfrozen variables versus
	the evolution time of {\tt SPINFLIP} (upper);
	the overlap histogram of $1000$ sampled solutions
	and the matrix of Hamming
	distances of these solutions for this formula (lower left)
	and for its shuffled version (lower right).}
\end{figure}

\begin{figure}[t]
\includegraphics[width=0.45\linewidth]{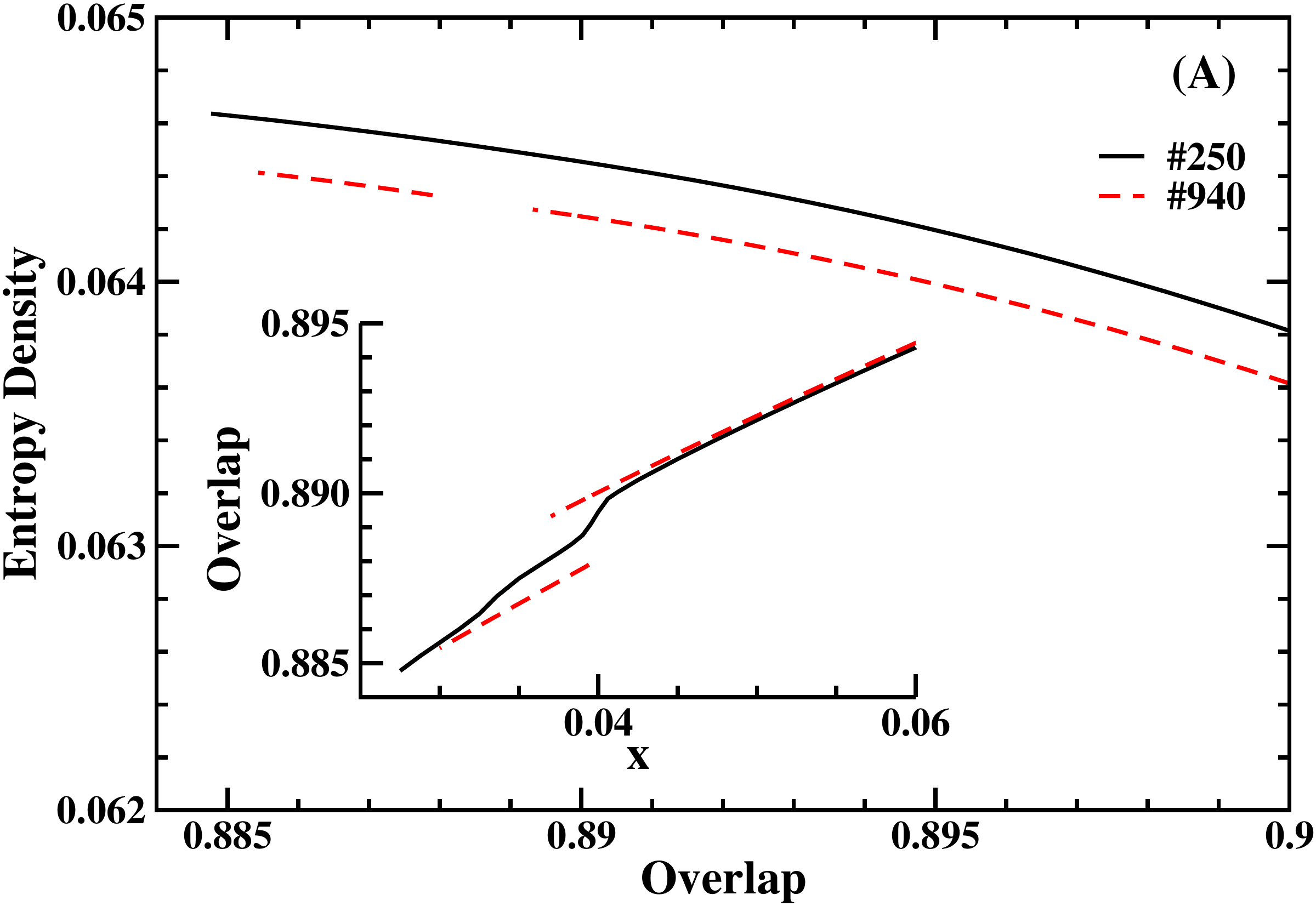}
\includegraphics[width=0.45\linewidth]{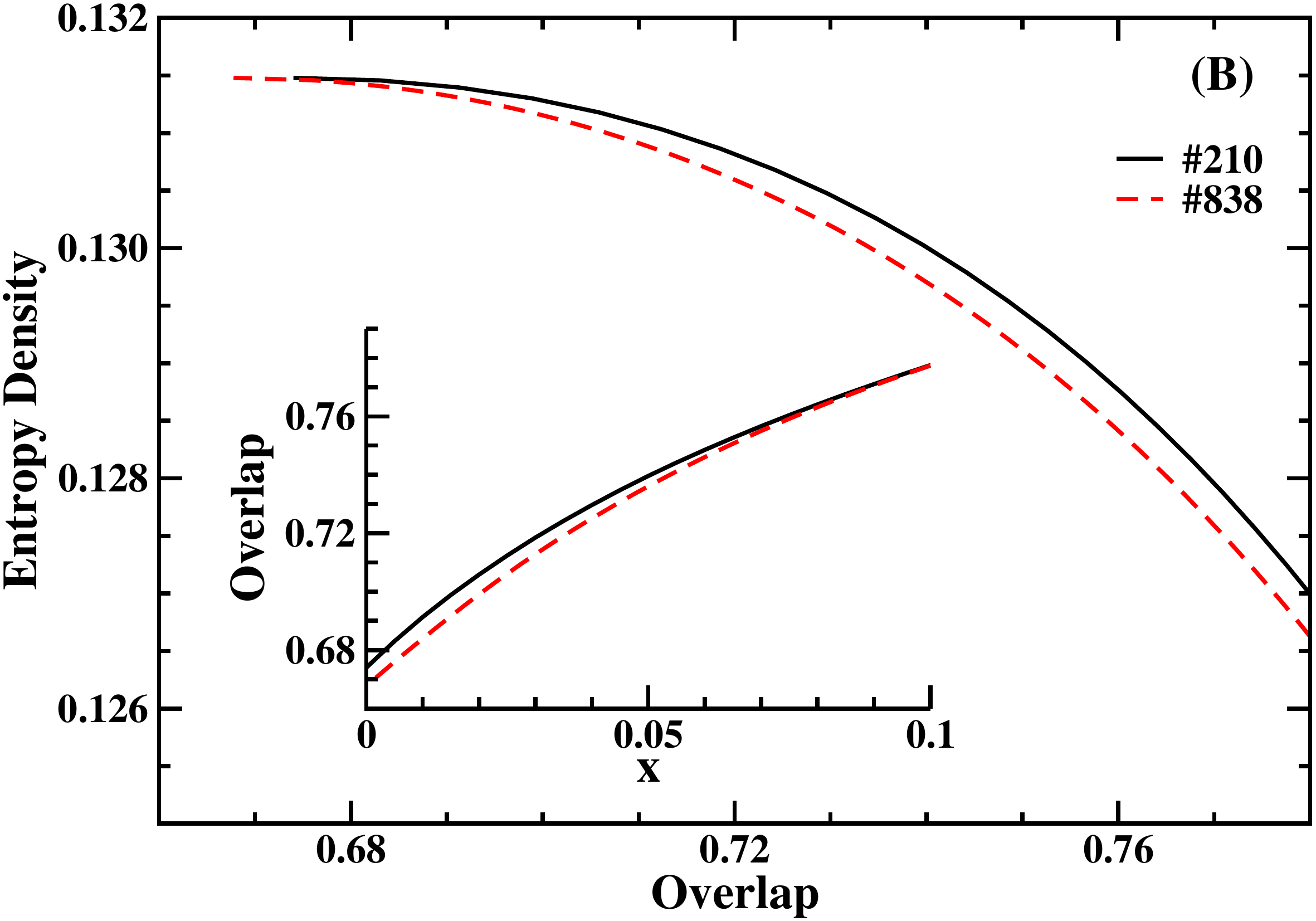}
\caption{\label{fig:entropya4p25}
	(Color online) 
	The entropy density $s(q)$ at a given overlap value $q$ with
	a reference solution. (A) Results for
	two solutions S-$250$ and S-$940$ of the
	lower left system of Fig.~\ref{fig:3sat4p25n1m}. 
	(B) Results for two solutions S-$210$ and S-$838$ of
	the lower right system of Fig.~\ref{fig:3sat4p25n1m}.
	The inset of (A) and (B) shows the overlap value $q$ as a
	function of the reweighting parameter $x$ of the replica-symmetric
	cavity method.
}
\end{figure}

As a first example, Fig.~\ref{fig:3sat4p25n1m} shows
the simulation results for a
random $3$-SAT formula of $N=10^6$. The constraint
density $\alpha=4.25$ of this formula is very close to the
satisfiability threshold $\alpha_s=4.267$, and the
initial solution $\vec{\sigma}^*$ for the {\tt SPINFLIP} random
walk process was obtained
by survey propagation \cite{Mezard-etal-2002}. The solid line in the
upper panel of Fig.~\ref{fig:3sat4p25n1m} is the number of
accumulated unfrozen variables $N_{\rm u}(t)\equiv |U(t)|$.
We notice that this number increases only slowly
(almost logarithmically) with evolution
time $t$, $N_{\rm u}(t) \sim \ln(t)$, and only $25\%$ of the variables are
found to be unfrozen at time $t=10^6$. The lower left panel of
Fig.~\ref{fig:3sat4p25n1m} is the
overlap histogram and the matrix of Hamming distances of $1000$ sampled
solutions
(with equal interval of $\Delta t= 1000$). 
As indicated by the fact that only a quarter of the variables have
been touched, the random walk process probably has visited only a
small fraction of the whole solution cluster in the relatively short 
evolution time of $10^6$. However, the overlap histogram and the
Hamming distance matrix 
clearly demonstrate that
the explored portion of the solution cluster is far from being
homogeneous. The overlap  histogram
has several peaks, and the Hamming distance matrix  shows that 
the sampled solutions can be divided into two
large groups, each of which can be further divided into
several sub-groups. The overlap of the visited
solutions with the initial solution $\vec{\sigma}^*$ has several
sudden drops as a function of $\ln t$, and each of these drops
is preceded by a plateau of overlap value (data not shown).
All these simulation results
are consistent with the proposal that several solution communities
exist in the studied solution cluster. The solutions of each community
are more densely connected to each other than to the outsider solutions.
Because of the dominance of intra-community connections in each
solution community, 
a random walker in a community-rich graph will be trapped in a single
community for a long time before it jumps into another community
and discovers new unfrozen variables. This proposed multi-trap mechanism
may be the reason of the logarithmic increase of
$N_{\rm u}(t)$ \cite{Bouchaud-Dean-1995}.

Guided by the Hamming distance matrix of Fig.~\ref{fig:3sat4p25n1m}
(lower left), we choose two sampled solutions,
solution S-$250$ and S-$940$ for entropy calculations
\cite{note3}. The overlap
between S-$250$ and S-$940$ is $0.8681$, and they are
suggested by Fig.~\ref{fig:3sat4p25n1m} (left lower) as belonging to
two different communities.
For S-$250$, the BP iteration is convergent as
long as the reweighting
parameter $x$ is in the range of $x\geq 0.0275$
(see Fig.~\ref{fig:entropya4p25}a). At $x=0.0275$, BP reports an
entropy density $s=0.06464$ and an overlap value $q = 0.8848$ with
S-$250$.
The overlap as a function of
$x$ has a rapid change at $x\approx 0.04$ (the same behavior is
observed for the entropy density), indicating a rapid change of
the statistical property of the solution cluster at $q\approx
0.890$ as viewed from S-$250$.
For S-$940$, BP is convergent when $x\geq 0.03$; at $x=0.03$
the entropy density is $s=0.06441$, and the overlap value is
$q=0.8854$. Two fixed-points of BP are obtained at
$0.035< x < 0.04$ for S-$940$ (Fig.~\ref{fig:entropya4p25}a),
indicating that there is a well-formed community of solutions whose
mean overlap with S-$940$ is $q\approx 0.890$, and this
community is embedded in a larger community of mean overlap
$q\approx 0.887$ with S-$940$.

The same numerical experiment is also carried out for
a random $3$-SAT formula of
$N=10^6$ and $\alpha=4.20$, starting from an initial solution obtained by
{\tt WALKSAT} \cite{Selman-Kautz-Cohen-1996,Seitz-Alava-Orponen-2005},
 and a set of random $3$-SAT formulas of
$N=20,000$ and $\alpha \in [3.825, 3.925]$, using initial solutions obtained by
belief propagation decimation
(see the following subsection) \cite{Krzakala-etal-PNAS-2007}.
The results of these
simulations suggest that the existence of community structure in single solution clusters
is a general property of random $3$-SAT formulas.

Given a solution $\vec{\sigma}^*$ for a formula $F$, we can shuffle
the connection pattern of $F$ to produce a maximally randomized
formula $F^\prime$ under the constraints that
(i) $\vec{\sigma}^*$ is still a solution of $F^\prime$,
(ii) each variable $i$ participates in the
same number of clauses as in $F$ and its spin value $\sigma_i^*$ satisfies
the same number of clauses as in $F$,
and (iii) each clause $a$ is satisfied by the same number of
spins of $\vec{\sigma}^*$ as in $F$.
When we run {\tt SPINFLIP} starting from $\vec{\sigma}^*$ for the shuffled
formula we are
unable to detect any community structures.  For the $3$-SAT formula of
$\alpha=4.25$
studied above, the simulation results obtained on a shuffled formula are
also shown in Fig.~\ref{fig:3sat4p25n1m}. The number $N_{\rm u}(t)$ of
discovered
unfrozen variables for this shuffled system has a sigmoid form as a function
of $\ln(t)$ and it
already reaches a high value of $0.9 N$ at time $t \sim 10^4$.
The overlap histogram of the
$1000$ sampled solutions (time interval $\Delta t=1000$)
has a  Gaussian form,
and the Hamming distance matrix of these
sampled solutions is featureless.

 This and additional
shuffling experiments confirm that community structure is present only in a
solution cluster of a random $3$-SAT formula but not in that of a 
shuffled formula.
The entropy calculations further confirms this point. For the randomized 
graph of Fig.~\ref{fig:3sat4p25n1m} (lower right), we have chosen two
most separated solutions S-$210$ and S-$838$ (with an overlap value
$0.6641$) to perform the entropy calculations.
The BP iteration is able to converge even when the reweighting
parameter decreases to zero, and at $x=0$ the same entropy density value
of $0.13148$ is reached (see Fig.~\ref{fig:entropya4p25}b). The overlap
$q$ as a function of $x$ does not show any signal of discontinuous
behavior.

\subsection{Community structures form before the clustering transition
	in random $3$-SAT}
\label{sec:3satn20k}

\begin{figure}[t]
	\includegraphics[width=0.9\linewidth]{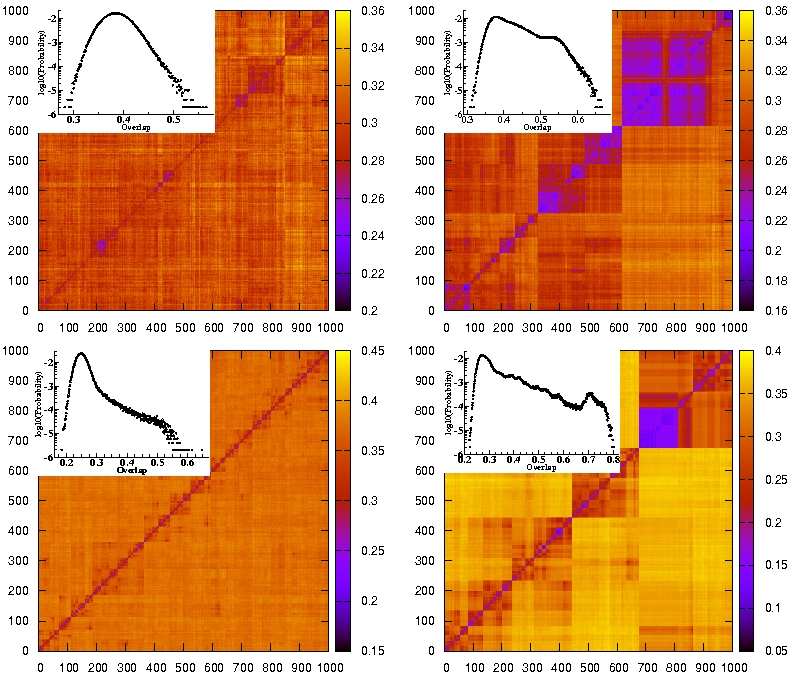}
	\caption{\label{fig:transition}
	 (Color online) The overlap histogram (in semi-logarithmic plot)
	and the matrix of Hamming
	distances of $1000$ sampled solutions for a random $K$-SAT formula
	of $20,000$ variables. {\tt SPINFLIP} first runs for $3\times 10^7$
	steps starting from a solution obtained by belief propagation
	decimation. Solutions are then
	sampled at equal time interval of $50,000$. The upper
	panel corresponds to $K=3$, $\alpha=3.825$ (left) and $\alpha=3.925$
	(right); the lower panel corresponds to  $K=4$ $\alpha=9.10$ (left)
	and $\alpha=9.22$ (right). The most probable overlap values
	in the shown overlap histograms of $\alpha=3.825$ ($K=3$),
	$\alpha=9.10$ and $9.22$ ($K=4$) are in agreement with the
	mean overlap values predicted by the replica-symmetric cavity
	method for the same formulas, indicating that the solution space
	for these formulas is composed of one single giant component.
	}
\end{figure}

\begin{figure}
\includegraphics[width=0.45\linewidth]{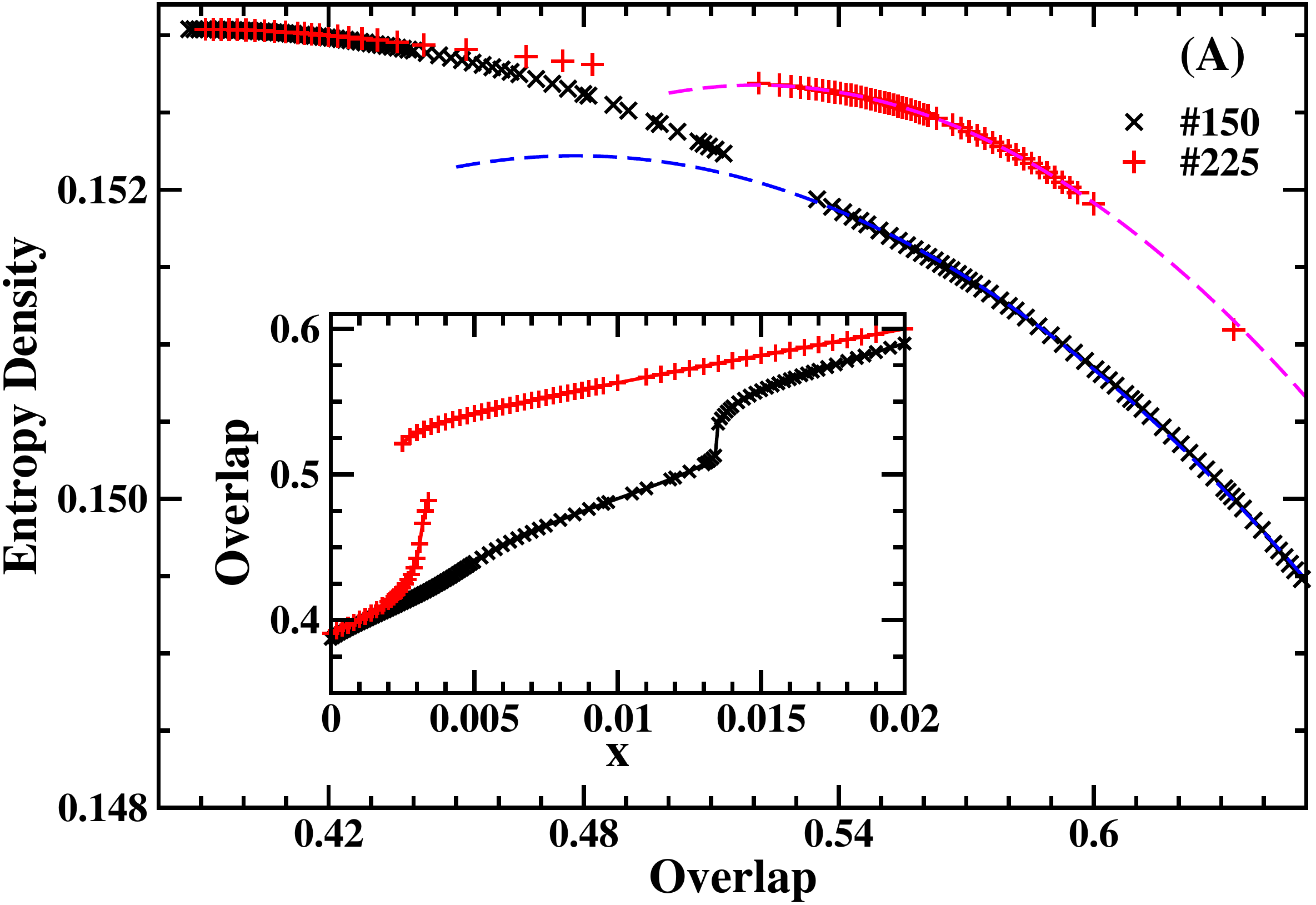}
\includegraphics[width=0.43\linewidth]{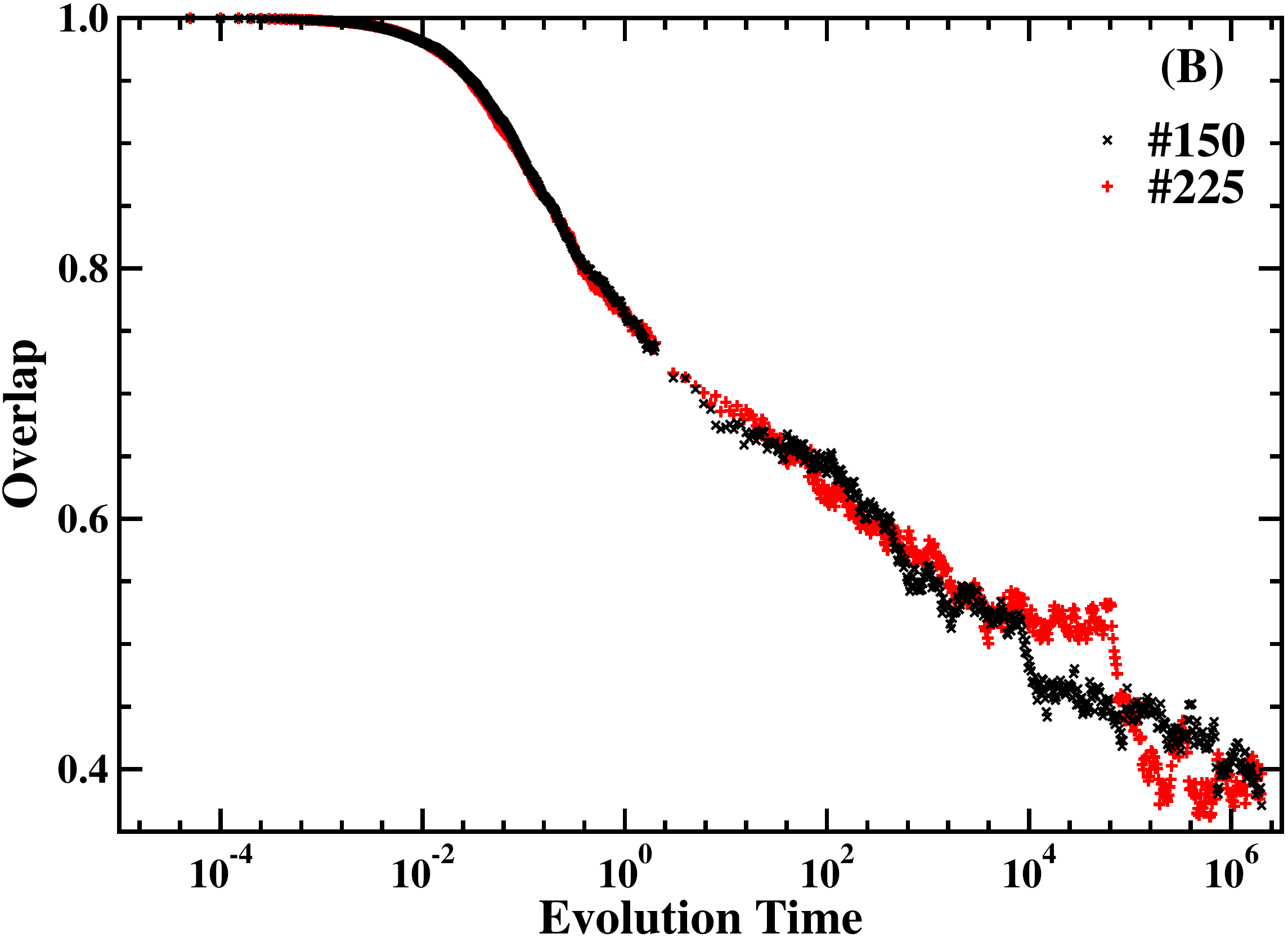}
\caption{\label{fig:a3p825} 
	(Color online) Structure of the solution cluster
	examined in Fig.~\ref{fig:transition} (upper left,
	$\alpha=3.825$).
	(A) The entropy density $s(q)$ of solutions at a given overlap value
	$q$ with reference solution S-$150$ and S-$225$.
	(B) Two overlap evolution trajectories starting
	from S-$150$ and S-$225$. An evolution trajectory is obtained by
	an unbiased random walk starting from either S-$150$ or S-$225$,
	the overlap of the visited solution with the starting solution
	is recorded during the random walk process.
	In (A) the two dashed lines are  fitting curves of the
	quadratic form $s(q)=s_0 - a_0 (q-q_0)^2$. The fitting
	parameters are $s_0=0.15222\pm 3\times 10^{-5}$,
	$q_0=0.478\pm 0.003$ (fitting range being
	$0.535\leq q \leq 0.65$, for S-$150$) and
	$s_0=0.153035\pm 1\times 10^{-6}$, $q_0=0.3900\pm 0.0004$
	($0.52\leq q\leq 0.6$, for S-$225$). The inset of (A) shows the
	overlap value $q$ as a function of the reweighting parameter $x$.
}
\end{figure}

\begin{figure}
\includegraphics[width=0.45\linewidth]{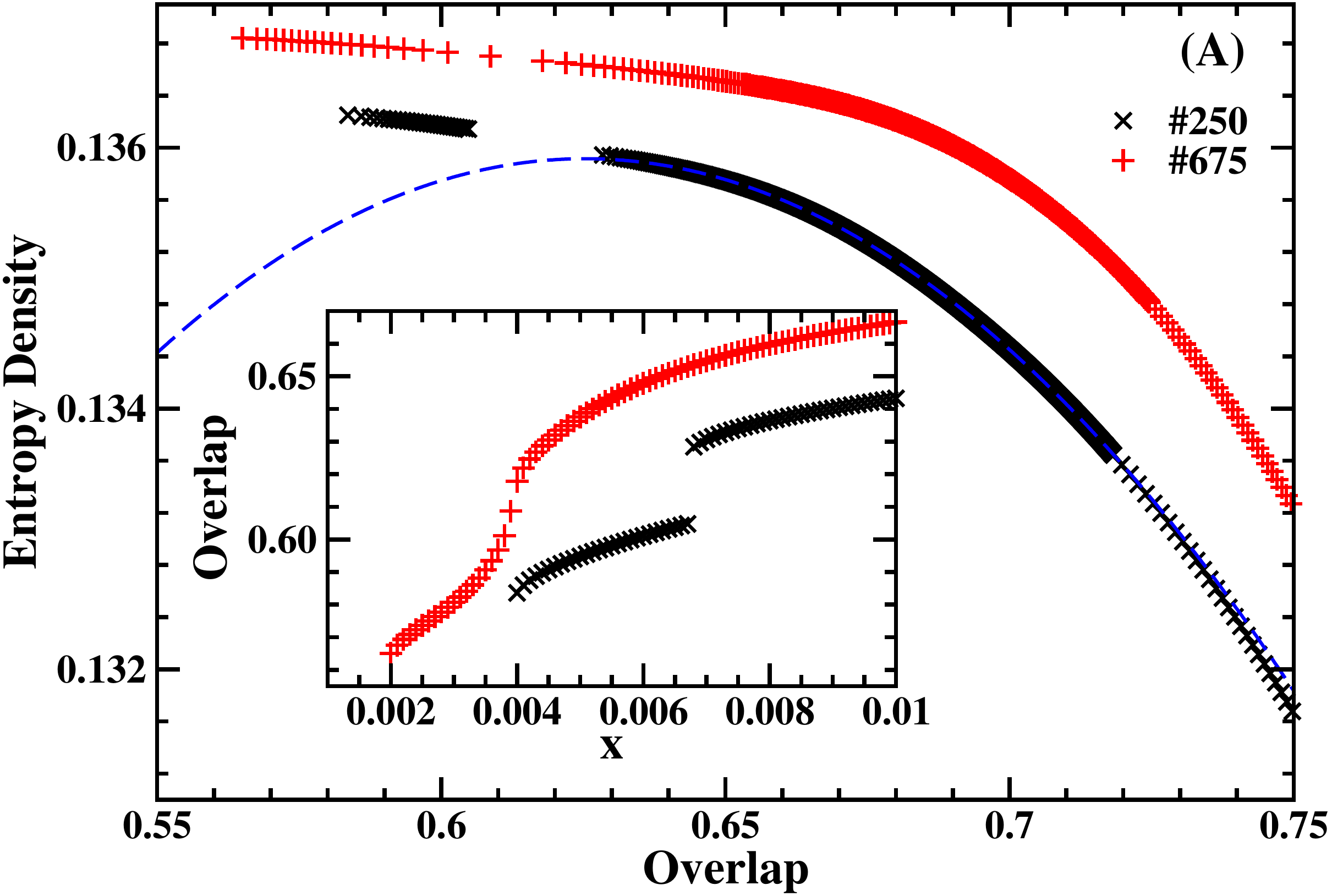}
\includegraphics[width=0.43\linewidth]{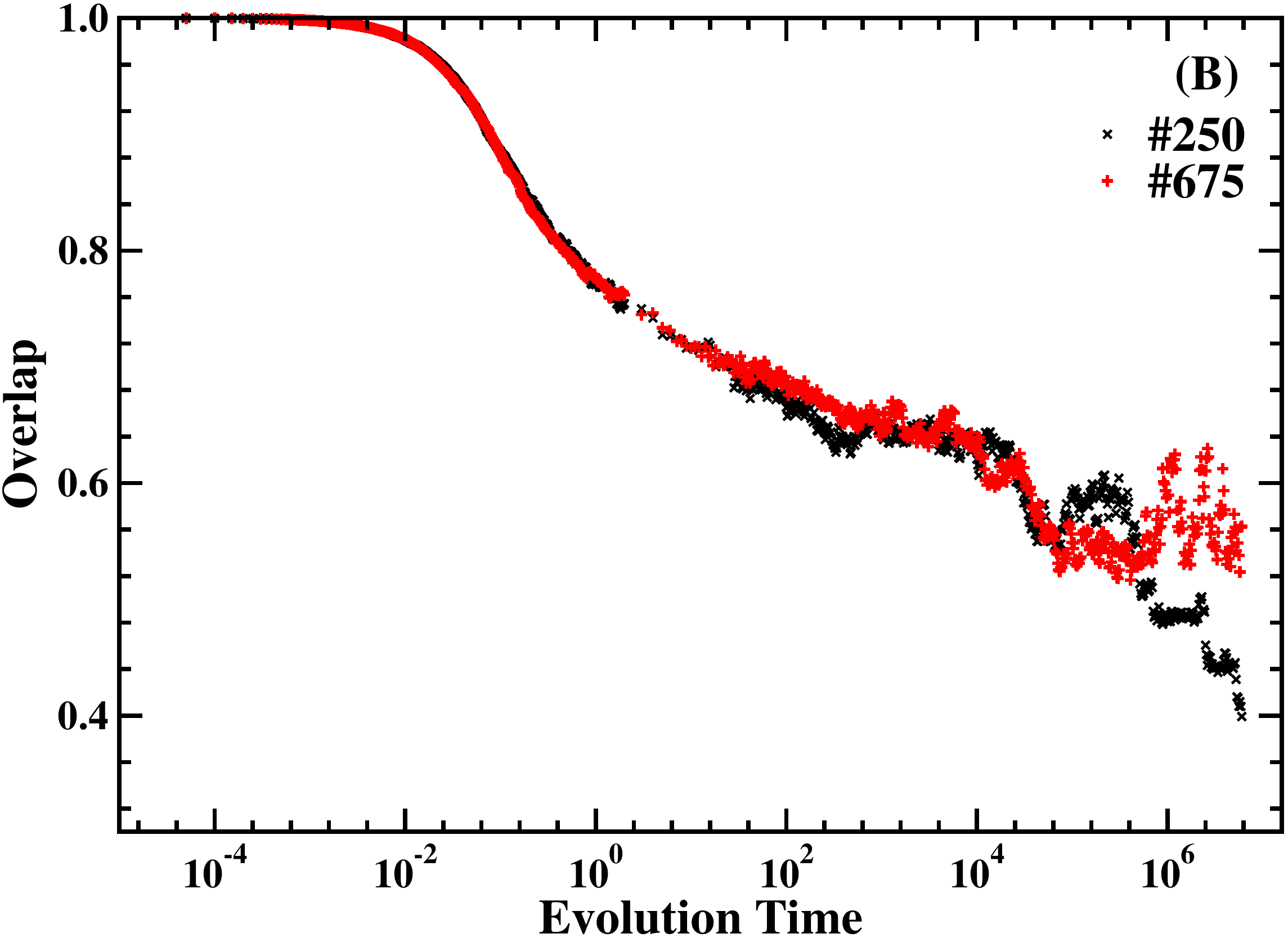}
\caption{\label{fig:a3p925}
	(Color online)
	Same as Fig.~\ref{fig:a3p825}, but the solution cluster is
	the one studied in Fig.~\ref{fig:transition} (upper right), with
	$\alpha=3.925$. The dashed curve in (A) 
	is a quadratic fitting curve $s(q)=s_0 - a_0 (q-q_0)^2$
	with fitting parameters $s_0=0.135916$,
	$q_0=0.625297\pm 7\times 10^{-5}$ (for
	S-$250$, fitting range being
	$0.628 \leq q \leq 0.72$).
}
\end{figure}
Krzakala {\em et al.} \cite{Krzakala-etal-PNAS-2007} predicted that a clustering
transition occurs in the solution space of a random $3$-SAT formula at 
the critical constraint density  $\alpha_d = 3.87$.
At this point, exponentially many Gibbs states
emerge in the solution space, with a few of these states dominating the
solution space. A Gibbs state of the mean-field statistical physics theory
is defined mainly in terms of the correlation property of the solution space.
It is regarded as a set of solutions within which there are no
long-range point-to-set correlations \cite{Montanari-Semerjian-2006}.
For a large random $K$-SAT formula,
whether there is a one-to-one correspondence between a solution
cluster (which is defined as a connected component of the solution space)
and a Gibbs state of statistical physics is still an open
question. But even if there is not a strict one-to-one correspondence,
it is natural to believe that a solution cluster and a Gibbs state of
solutions are closely related. In this section, we investigate the structure
of a single solution cluster of a random $3$-SAT formula at $\alpha$ close
to $\alpha_d$ by extensive {\tt SPINFLIP} simulations
on random $3$-SAT formulas of size $N=20,000$.
Ten random $3$-SAT formulas are generated at each of
the constraint density values $\alpha \in \{3.825, 3.85, 3.875, 3.90, 3.925\}$,
and for each of these formulas a solution $\vec{\sigma}^*$ is
constructed using belief propagation decimation
\cite{Krzakala-etal-PNAS-2007}, which is then used by
{\tt SPINFLIP} as the starting point.

The belief propagation decimation program fixes variables of the input formula
sequentially with an interval of at least $50$ iterations, and it assigns a
spin value to a variable according the predicted marginal spin distribution.
We have chosen such an extremely slow fixing protocol with the hope of being
able to pick a solution uniformly random from the solution space. 
For $\alpha = 3.825$ and $3.85$, we are able to calculate 
the entropy density of the whole solution space of a formula
and the mean overlap between two solutions
using the replica-symmetric cavity method, with all the cavity fields
initially setting to zero
\cite{Li-Ma-Zhou-2009}. We have verified that the mean overlap
and entropy density values of the solution clusters
explored by {\tt SPINFLIP} are in agreement with the statistical
physics predictions. This is consistent with the belief
that the whole solution space is ergodic and has only a single (statistically
relevant) solution cluster.
For $\alpha=3.875, 3.90, 3.925$, the replica-symmetric cavity method
no longer converges on a single formula, and therefore we are not sure
whether the explored solution clusters are the dominating clusters. This
later ambiguity may not be too significant, as we are mainly interested in
the property of the solution cluster before the clustering transition.

In each run of {\tt SPINFLIP}, the random walk first runs at least
$3\times 10^7$ time steps starting from the input solution, and then
$1000$ solutions are sampled at equal time interval of $\Delta t = 50,000$.
Before sampling of solutions, {\tt SPINFLIP} has enough to time to
flip almost all the variables, therefore during the later solution
sampling process, {\tt SPINFLIP} actually performs an unbiased random walk.

The overlap histograms and Hamming distance matrices of the
sampled solutions at $\alpha = 3.825$ show only
weak heterogeneous features
(a typical example is shown in Fig.~\ref{fig:transition} upper left);
but as  $\alpha$ increases, the heterogeneity of the solution
cluster becomes more and more evident 
(for $\alpha = 3.925$, a typical
example is shown in Fig.~\ref{fig:transition} upper right). These
results might indicate that only weak community structure is present
in the studied solution clusters of $\alpha=3.825$. However,
we must be careful to draw conclusions from figures such as
Fig.~\ref{fig:transition}, as the community structures revealed by
{\tt SPINFLIP} also depend on the time interval
$\Delta t$ of solution sampling. Even if the solution cluster
is composed of extremely many
communities, if $\Delta t$ is of the same order as 
the typical trapping times of the communities, two sampled
solutions of {\tt SPINFLIP} will only have a low probability of belonging
to the same community. Then the Hamming distance matrix of the
sampled solutions will be very homogeneous.
For the case of Fig.~\ref{fig:transition} (upper left), we find that
$\Delta t = 50,000$ is comparable to the
typical trapping time of a community (see Fig.~\ref{fig:a3p825}b).
If $\Delta t$ is chosen to be ten times shorter,
the sampled solutions show very evident community structures
also at $\alpha = 3.825$ (data not shown).

The clustering analysis of sampled solutions is complemented by
entropy calculations. For the example of $\alpha=3.825$ shown
in Fig.~\ref{fig:transition} (upper left),
we have calculated the entropy densities of
solutions at a given overlap with two reference solutions S-$150$ and
S-$225$. The results are shown in Fig.~\ref{fig:a3p825}.
For solution S-$150$, as the reweighting parameter $x$
decreases to $x=0.0135$,
both the entropy density and the overlap show a sudden change.
This behavior indicates that S-$150$ is contained in a solution
community of entropy
density $s \approx 0.1519$ and of mean overlap
$q\approx 0.5349$ with S-$150$. 
On the other hand, the whole solution cluster has an
entropy density $s =0.15304$ and mean overlap $q=0.3872$ with
S-$150$.
We have performed an unbiased random walk simulation
starting from S-$150$ (see Fig.~\ref{fig:a3p825}b) to find that
the overlap as a function of evolution time (in logarithmic scale)
indeed has an evident plateau at  $q\approx 0.53$ before it eventually
decays to $q\approx 0.39$.

For the solution S-$225$,  Fig.~\ref{fig:a3p825}a
shows that there is a region of
the reweighting parameter $x$ within which two fixed-point solutions of the
BP iteration equations coexist. One of the fixed-point of BP describes
the statistical property of the solution community,
which has an entropy density
$s\approx 0.1527$ and mean overlap $q\approx 0.5212$ with
S-$225$, while the other fixed-point describes the statistical
property of the whole solution cluster, which has an entropy
density $s=0.15304$ and mean overlap $q=0.3911$ with
S-$225$. If we perform an unbiased random walk process in the solution
cluster starting from solution S-$225$, we find that the overlap with
S-$225$ stays at a plateau value of $q\approx 0.52$ for a long time until
it suddenly (in logarithmic scale) drops to a value of $q\approx 0.39$
(see Fig.~\ref{fig:a3p825}b), in agreement with the replica-symmetric 
BP results. Similar results are obtained from other sampled solutions.

From the different entropy density values of the communities and the
fact that the two reference solutions S-$150$ and S-$225$  have
a small overlap of $q\approx 0.39$, we conclude that they belong to
different communities of the same solution cluster. And from the fact that the
entropy density of the examined solution cluster is the same as the entropy
density of the whole solution space (the
later is obtained by the replica-symmetric
BP with both random and zero initial conditions \cite{Li-Ma-Zhou-2009}),
we conclude this solution cluster is actually the only statistically relevant
solution cluster of the whole solution space.
Qualitatively the same results
are obtained for the other studied random $3$-SAT formulas of $\alpha=3.825$
and $\alpha=3.85$.
We therefore conclude that many solution communities have already formed
in the single statistically relevant solution cluster of a large random
$3$-SAT formula at constraint density $\alpha < \alpha_d$. If the solution
cluster breaks into many connected components
at the clustering transition point $\alpha_d$, this ergodicity breaking
can be understood as the final separation of groups of communities caused
by the loss of inter-community links.

When the constraint density $\alpha$ is beyond the clustering transition
value $\alpha_d$, all the explored single solution communities of the
random $3$-SAT formulas demonstrate clear community structures,
according to the overlap histogram and Hamming distance matrices of the
sampled solutions (see Fig.~\ref{fig:transition} upper right for a
typical example). 
The existence of community structure in single solution clusters is
also confirmed by entropy calculations. As an example, we show in
Fig.~\ref{fig:a3p925}a the results of the replica-symmetric cavity
method on a solution cluster that corresponds to Fig.~\ref{fig:transition}
upper right ($\alpha=3.925$). We choose solution S-$250$ and S-$675$ 
(with mutual overlap $0.3842$) as
two reference solutions (similar results are obtained for other sampled
solutions). For S-$250$, the entropy density and overlap
value change suddenly when the reweighting parameter is decreased to
$x=0.0068$, indicating that S-$250$ belongs to a
solution community of entropy density $s \approx 0.13595$
and mean overlap $q\approx 0.628$ with S-$250$.
This solution community is itself contained in a larger
community of entropy density $s \geq 0.13625$ and mean overlap
$q\leq 0.5835$ with S-$250$. The evolution trajectory of the overlap
value with S-$250$ as obtained from an unbiased random walk process
(Fig.~\ref{fig:a3p925}b), which
has a series of plateaus of decreasing heights, is consistent with such a
nested (hierarchical) organization of communities. For
S-$675$, the entropy data suggest that it belongs to
a different community of entropy density $s \approx 0.1367$, whose
mean overlap with S-$675$ is $q \approx 0.62$. This solution community
itself form a subgraph of a larger community of entropy density
$s\geq 0.13685$ and of mean overlap $q\leq 0.565$ with S-$675$.
The overlap evolution trajectory starting from S-$675$ jumps
between the values of $q\approx 0.63$ and $q\approx 0.53$ at $t\geq 10^6$.
This jumping behavior demonstrates that the unbiased random walker is able to
visit the solution community of S-$675$ frequently. This probably indicates
that the community of S-$675$ is one of the largest communities of the solution
cluster.

For the studied solution cluster at $\alpha=3.925$,
when the reweighting parameter $x$ is very small ($x<0.004$ for S-$250$ and
$x<0.002$ for S-$675$), we are unable to find a fixed-point for
the replica-symmetric BP equations. As $x$ approaches
zero, the corresponding dominating solutions probably are distributed into
different solution clusters, and the replica-symmetric cavity method is
no longer sufficient to describe their statistical properties.

\section{Results for random $4$-SAT formulas}
\label{sec:4sat}

\subsection{Results for a large random $4$-SAT formula with $\alpha > \alpha_d$}

\begin{figure}[t]
	\includegraphics[width=0.9\linewidth]{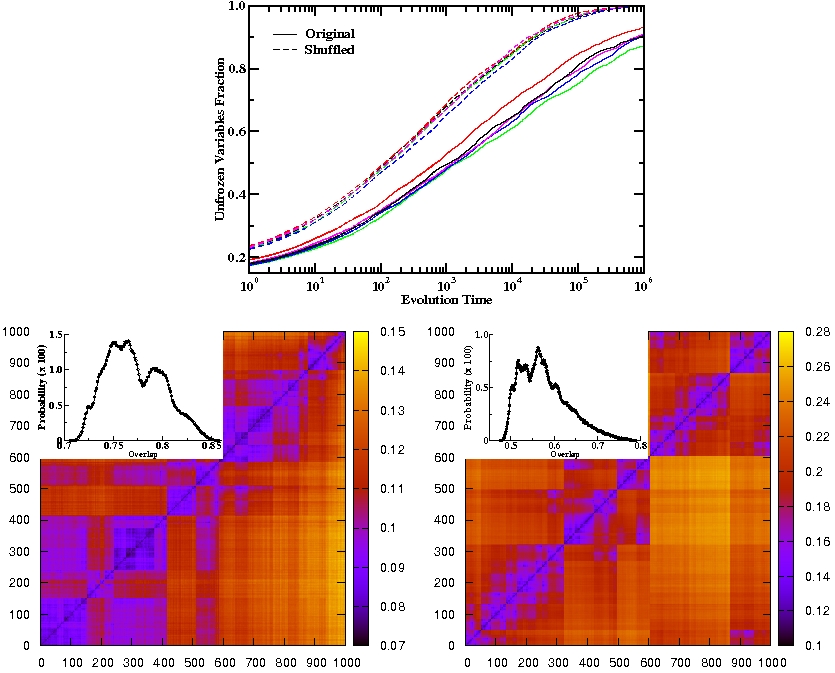}	
	\caption{\label{fig:4sat9p46n100k}
	(Color online) Simulation results on a random $4$-SAT
	formula with $N=10^5$ variables and constraint density
	$\alpha=9.46$. (upper) Number of discovered unfrozen variables versus
	the evolution time of {\tt SPINFLIP}, starting from five
	different initial solutions. (lower left and lower right)
	The overlap histogram of $1000$ sampled solutions from one
	initial solution and the matrix of Hamming
	distances of these solutions for this formula (lower left)
	and its shuffled version (lower right).}
\end{figure}

\begin{figure}
\includegraphics[width=0.45\linewidth]{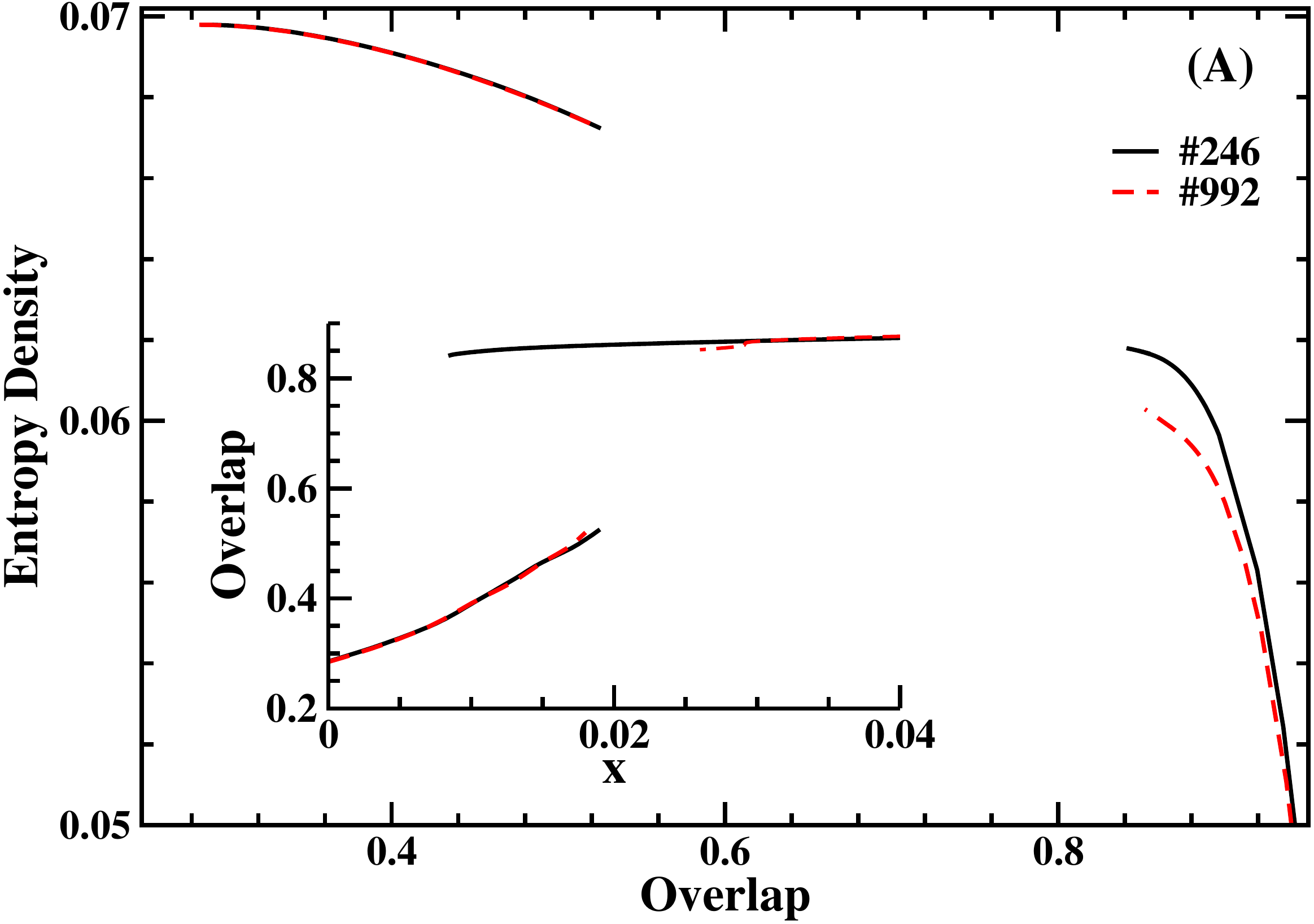}
\includegraphics[width=0.45\linewidth]{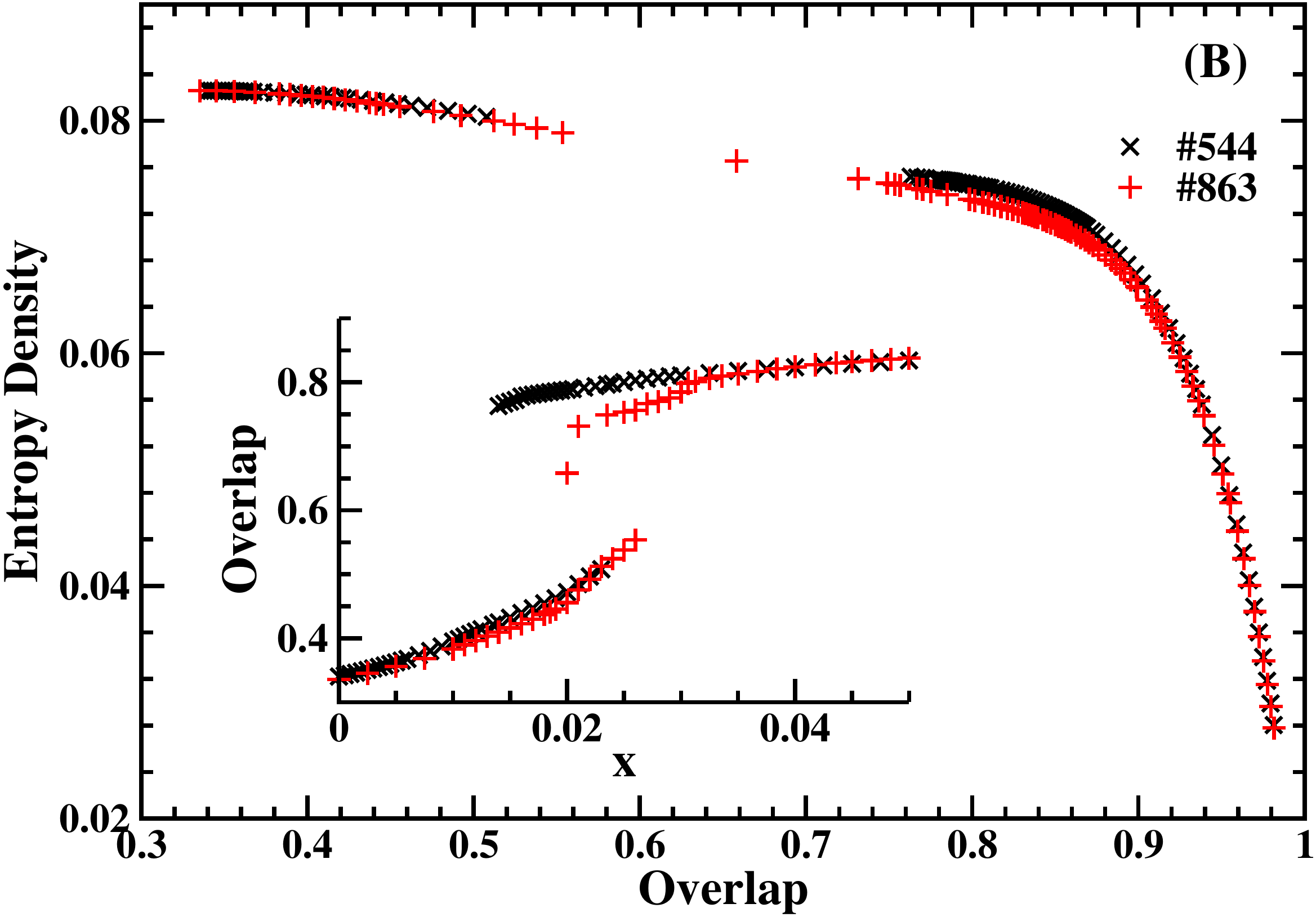}
\caption{\label{fig:entropya9p46n100k}
(Color online)
 The entropy density curves $s(q)$ as a function of
overlap $q$. (A) Results obtained by choosing two reference
solutions S-$246$ and S-$992$ in the solution cluster
of Fig.~\ref{fig:4sat9p46n100k} (lower left).
(B) Results obtained by choosing two reference solutions S-$544$ and
S-$863$ in the solution cluster of Fig.~\ref{fig:4sat9p46n100k} (lower right).
The inset in each sub-figure is the overlap value $q$ as a function of the
reweighting parameter $x$.
}
\end{figure}

We perform simulations on a single large random $4$-SAT formula $F$ of
$N=10^5$ variables. The constraint density of the formula is
$\alpha=9.46$, beyond the clustering transition point 
$\alpha_d=9.38$ \cite{Krzakala-etal-PNAS-2007}.
Five solutions were
obtained using belief propagation decimation
 for this formula; $F$ was then shuffled with
respect to each of these solutions to obtain
five new formulas $F^\prime$
(see Sec.~\ref{sec:3sat4p25n1m}).
The number $N_{\rm u}(t)$ of discovered unfrozen variables
as a function of the evolution time $t$ of {\tt SPINFLIP}
on these ten instances are shown in
Fig.~\ref{fig:4sat9p46n100k} (upper panel). There is no qualitative
difference between the $N_{\rm u}(t)$ curves of the original formula
and those of the shuffled formulas, as compared with the
results of the random $3$-SAT case in Fig.~\ref{fig:3sat4p25n1m}.
The random walk process is
able to flip most of the variables at least once
in an evolution time of $t=10^6$ both on the original and on the shuffled
formulas.

The lower left and lower right panel of Fig.~\ref{fig:4sat9p46n100k} are,
respectively, the overlap histogram and Hamming distance matrix of
$1000$ sampled solutions at time interval $\Delta t=1000$ for the original
formula and one of its shuffled version, with the random walk process
starting from the same initial solution. From these two figures, we
infer that both the solution cluster of
the original and the shuffled formula have non-trivial
community structures. This is another important difference 
compared with the random $3$-SAT results shown in Fig.~\ref{fig:3sat4p25n1m},
where the solution cluster of the shuffled formula does not
show community structure.

For the solution cluster of Fig.~\ref{fig:4sat9p46n100k} (lower left), we
choose two solutions S-$246$ and S-$992$ (with an overlap of
$0.7064$) for entropy calculations.  The entropy density curves $s(q)$ as
a function of the overlap $q$ with these two solutions are
shown in Fig.~\ref{fig:entropya9p46n100k}a.
For S-$246$, the replica-symmetric BP iteration equations have
two fixed points when the reweighting parameter is in the range of
$0.0084 \leq x \leq 0.019$. The fixed point with $q>0.8$ corresponds
to the local solution community of S-$246$, which has an entropy density of
$s\approx 0.06180$ and mean overlap $q\approx 0.841$ with S-$246$.
The other fixed point with $q<0.56$ probably corresponds to the whole
solution space, which has an entropy density $s=0.069794$ at $x=0$.
For S-$992$, the BP iteration equations are convergent for $x\geq 0.026$ and
$x\leq 0.018$ but are divergent for $0.018 < x < 0.026$. We infer that
S-$992$ is associated with a solution community of entropy density
$s=0.0603$, whose mean overlap with S-$992$ is $q\approx 0.852$.
These entropy results confirm the indication of
Fig.~\ref{fig:4sat9p46n100k} (left lower) that S-$246$ and S-$992$
belong to two different communities (of the same cluster).
As the constraint density of the formula is beyond the
clustering transition point $\alpha_d$,
its solution space very probably is composed of many
extensively separated solution clusters. In agreement with this
expectation, the mean-field cavity method
predicts that the mean overlap of the whole solution space to
the explored solution cluster is $q\approx 0.3$,

For the solution cluster of the shuffled formula studied in
Fig.~\ref{fig:4sat9p46n100k} (lower right), we also choose two solutions
S-$544$ and
S-$863$ (with mutual overlap $0.47636$) for
entropy calculations. The
results shown in Fig.~\ref{fig:entropya9p46n100k}b confirm that
the solution cluster of the shuffled formula has different
communities. The community of S-$544$ has an entropy density
$s\approx 0.07518$ and a mean overlap $q\approx 0.7630$ with S-$544$,
while that of S-$863$ has an entropy density
$s\approx 0.07324$ and a mean overlap $q\approx 0.7985$ with S-$863$.
As indicated by the small breaks of the
$s(q)$ curve of S-$863$ in Fig.~\ref{fig:entropya9p46n100k}b, the local
community of S-$863$ probably is a sub-graph of a larger community of
entropy density $s\approx 0.0745$, whose mean overlap with S-$863$ is
$q\approx 0.75$.
The entropy density of the whole solution space
as obtained at $x=0$ is $s=0.0825785$. The mean overlap of the whole solution
space to either of the two reference solutions is $q \approx 0.34$.

\subsection{Community structures form before the clustering transition
	in random $4$-SAT}
\label{sec:4satn20k}

\begin{figure}
\includegraphics[width=0.45\linewidth]{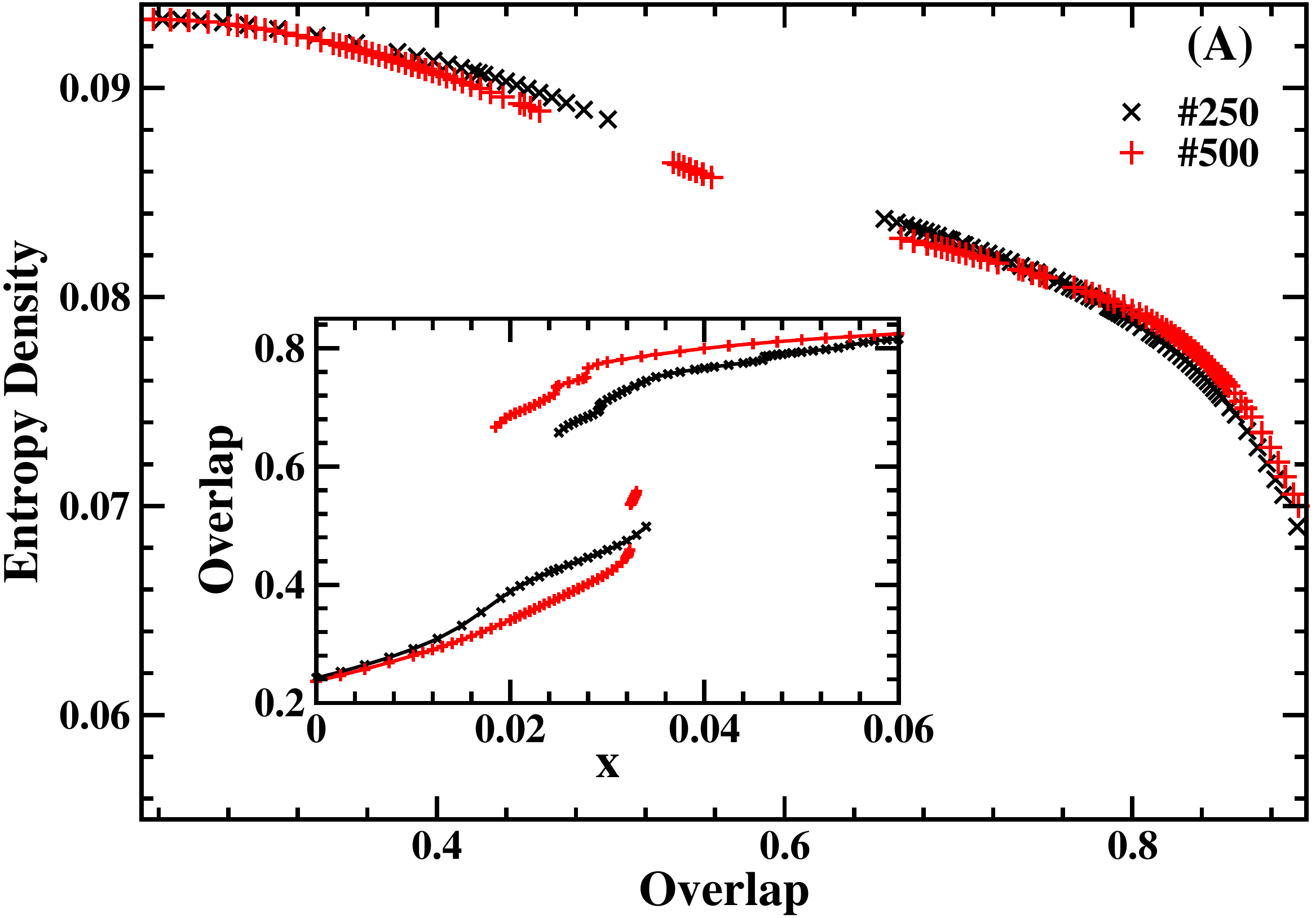}
\includegraphics[width=0.43\linewidth]{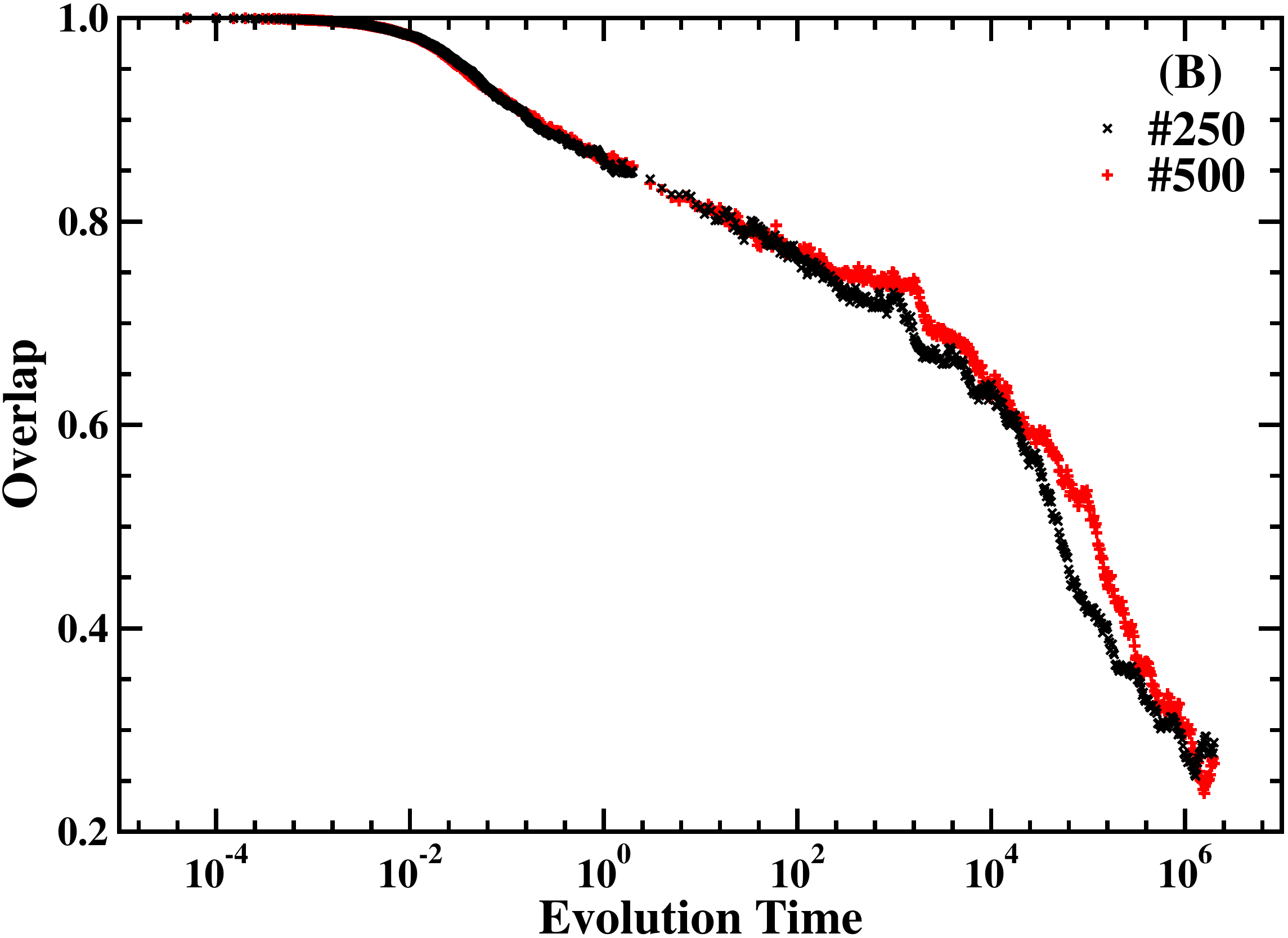}
\caption{\label{fig:a9p10}
	Same as Fig.~\ref{fig:a3p825}, but the solution cluster is
	for a $4$-SAT formula of $\alpha=9.10$, whose Hamming distance
	matrix is shown in the lower left panel of Fig.~\ref{fig:transition}.
}
\end{figure}

Similar to Sec.~\ref{sec:3satn20k}, we continue to
investigate whether solution communities have formed in the
solution space of a random $4$-SAT formula before the
clustering transition point $\alpha_d=9.38$.
For each of the constraint densities $\alpha
\in \{9.10, 9.22, 9.30, 9.38, 9.46, 9.54 \}$, ten random $4$-SAT formulas
of $N=20,000$ variables are generated, and a solution is obtained by
belief propagation decimation
 for each of these formulas. We then use the same
random walk protocol as mentioned in Sec.~\ref{sec:3satn20k} to
sample a large number of solutions for clustering analysis. Two
typical solution-clustering results, one for a formula
with $\alpha=9.10$ and the other for a formula with $\alpha=9.22$,
are shown in Fig.~\ref{fig:transition} lower left and lower right.

Our simulation results reveal that the connection patterns of
all these studied solution clusters at
$9.10\leq \alpha \leq 9.54$ are far from being homogeneous.
The lower panel of Fig.~\ref{fig:transition} indicates that there are
already many
small solution communities in the solution cluster of $\alpha=9.10$; and
that the community structures of the solution cluster will be more and
more pronounced as $\alpha$ increases. To be more quantitative, we have
calculated the statistical properties of solution communities by
performing BP iterations (with a reweighting
parameter $x$) starting from various sampled solutions.
We show as an example the results of the entropy
calculations performed on two solutions S-$250$ and S-$500$ of the
solution cluster of Fig.~\ref{fig:transition} (lower left), with
$\alpha=9.10$. Similar to what we have observed before, as the reweighting
parameter $x$ decreases, the entropy density and overlap values predicted by
the replica-symmetric cavity method show several small sudden changes,
and at $x \sim 0.03$ the BP equations have more than one
fixed-point solutions. From these results, we estimate that the
solution cluster that contains S-$250$ has an entropy density of
$s\approx 0.08374$ and a mean overlap $0.6574$ with S-$250$, while the
solution community of S-$500$ has an entropy density $s\approx 0.08281$
 and a mean overlap
$q\approx 0.6668$ with S-$500$.
Both of these two solution communities probably have non-trivial
internal structures, as indicated by the sudden small drops of
the overlap value as a function of $x$ (see the inset of
 Fig.~\ref{fig:a9p10}a).
The whole solution cluster has an entropy density
$s=0.093279$ and a mean overlap $q\approx 0.24$ with either of these two
reference solutions. These results are confirmed by the two overlap evolution
trajectories shown in Fig.~\ref{fig:a9p10}b, which show several
plateaus at $q\approx 0.7$ in the semi-logarithmic plot. The fact that
overlap values with S-$250$ and S-$500$ fluctuate at long times around
the theoretically predicted value of $q\approx 0.24$ confirms that the
studied solution cluster is the only statistically relevant cluster of the
whole solution space.

\section{Conclusion}
\label{sec:conclusion}

In summary, this work studied the solution space statistical properties
of large random random $3$- and $4$-SAT formulas by
extensive random walk simulations and by the replica-symmetric
cavity method of statistical physics. A solution space is mapped to 
a huge graph, in which each vertex represents an
individual solution and the edge between two vertices means that the
two corresponding solutions differ on just one variable. 
A solution cluster of the solution space is defined as a connected
component of solutions, and a solution community of a solution
cluster is a
set of solutions which are more similar with each
other and more densely inter-connected with each other than with
the outsider solutions of the solution cluster.
The results of this paper suggest that, as the constraint density
$\alpha$ of a random $K$-SAT ($K=3,4$) formula
increases, the solution space of the formula
first forms many solution communities
before the solution space experiences
a clustering transition at the critical constraint density
$\alpha_d$.
For $\alpha> \alpha_d$, the results of this paper also
suggests that  the individual solution clusters of
the solution space (which may correspond to
different solution Gibbs states) still have rich internal
community structures. The entropy density of a single solution community
in a solution cluster  is calculated by 
belief propagation iteration with a reweighting parameter $x$. From the
observed discontinuity of the overlap $q$ (with a given reference solution) at
certain critical values of $x$, we infer that the solution communities 
can be regarded as well-defined thermodynamic phases of the partition
function Eq.~(\ref{eq:partitionfunction}). 

As the constraint density $\alpha$ of a random $K$-SAT formula increases,
the density of inter-community connections in its solution space will
decrease. Therefore the solution space will split into many
solution clusters as $\alpha$ becomes large enough. Very probably
the splitting of the
solution space is not a gradual process, with the solution clusters
being divided from the single giant component one after another, but
rather being a highly cooperative process with
(exponentially) many solution clusters emerge at a critical constraint
density $\alpha_d^\prime$.
If this is really the case, it is very interesting to know
whether in the thermodynamic limit of $N\rightarrow \infty$ the
value of $\alpha_d^\prime$ is identical to $\alpha_d$.
One way to check this is to perform simulations on the solution space
using two mutually attractive random walkers
\cite{Ciliberti-Martin-Wagner-2007-b}). One may also
simultaneously follow the evolution
processes of many different solution communities of the same random
$K$-SAT formula as a function of the constraint density $\alpha$.

The main qualitative results of this paper are expected to be applicable
also to large  random $K$-SAT formula with $K\geq 5$. They may also
be applicable to other random constraint satisfaction problems such as
the random coloring problem.

We have not yet investigated the lowest value of $\alpha$ at which
solution communities begin to emerge in the solution space of a
random $K$-SAT formula. This is an important open question for future
studies.

\section*{Acknowledgement}

HZ thanks Silvio Franz and Marc M{\'{e}}zard for helpful discussions and
KITPC (Beijing), LPTMS (Orsay), NORDITA (Stockholm) for hospitality.
This work was partially supported by the NSFC (10774150) and the 
China 973-Program (2007CB935903). The computer simulations were performed
on the HPC cluster of ITP.

%\bibliography{/cygdrive/d/ResearchPapers/references}

\begin{thebibliography}{24}
\expandafter\ifx\csname natexlab\endcsname\relax\def\natexlab#1{#1}\fi
\expandafter\ifx\csname bibnamefont\endcsname\relax
  \def\bibnamefont#1{#1}\fi
\expandafter\ifx\csname bibfnamefont\endcsname\relax
  \def\bibfnamefont#1{#1}\fi
\expandafter\ifx\csname citenamefont\endcsname\relax
  \def\citenamefont#1{#1}\fi
\expandafter\ifx\csname url\endcsname\relax
  \def\url#1{\texttt{#1}}\fi
\expandafter\ifx\csname urlprefix\endcsname\relax\def\urlprefix{URL }\fi
\providecommand{\bibinfo}[2]{#2}
\providecommand{\eprint}[2][]{\url{#2}}

\bibitem[{\citenamefont{M{\'{e}}zard et~al.}(2002)\citenamefont{M{\'{e}}zard,
  Parisi, and Zecchina}}]{Mezard-etal-2002}
\bibinfo{author}{\bibfnamefont{M.}~\bibnamefont{M{\'{e}}zard}},
  \bibinfo{author}{\bibfnamefont{G.}~\bibnamefont{Parisi}}, \bibnamefont{and}
  \bibinfo{author}{\bibfnamefont{R.}~\bibnamefont{Zecchina}},
  \bibinfo{journal}{Science} \textbf{\bibinfo{volume}{297}},
  \bibinfo{pages}{812} (\bibinfo{year}{2002}).

\bibitem[{\citenamefont{Biroli et~al.}(2000)\citenamefont{Biroli, Monasson, and
  Weigt}}]{Biroli-Monasson-Weigt-2000}
\bibinfo{author}{\bibfnamefont{G.}~\bibnamefont{Biroli}},
  \bibinfo{author}{\bibfnamefont{R.}~\bibnamefont{Monasson}}, \bibnamefont{and}
  \bibinfo{author}{\bibfnamefont{M.}~\bibnamefont{Weigt}},
  \bibinfo{journal}{Eur. Phys. J. B} \textbf{\bibinfo{volume}{14}},
  \bibinfo{pages}{551} (\bibinfo{year}{2000}).

\bibitem[{\citenamefont{Achlioptas et~al.}(2005)\citenamefont{Achlioptas, Naor,
  and Peres}}]{Achlioptas-Naor-Peres-2005}
\bibinfo{author}{\bibfnamefont{D.}~\bibnamefont{Achlioptas}},
  \bibinfo{author}{\bibfnamefont{A.}~\bibnamefont{Naor}}, \bibnamefont{and}
  \bibinfo{author}{\bibfnamefont{Y.}~\bibnamefont{Peres}},
  \bibinfo{journal}{Nature} \textbf{\bibinfo{volume}{435}},
  \bibinfo{pages}{759} (\bibinfo{year}{2005}).

\bibitem[{\citenamefont{Krzakala et~al.}(2007)\citenamefont{Krzakala,
  Montanari, {Ricci-Tersenghi}, Semerjian, and
  Zdeborova}}]{Krzakala-etal-PNAS-2007}
\bibinfo{author}{\bibfnamefont{F.}~\bibnamefont{Krzakala}},
  \bibinfo{author}{\bibfnamefont{A.}~\bibnamefont{Montanari}},
  \bibinfo{author}{\bibfnamefont{F.}~\bibnamefont{{Ricci-Tersenghi}}},
  \bibinfo{author}{\bibfnamefont{G.}~\bibnamefont{Semerjian}},
  \bibnamefont{and}
  \bibinfo{author}{\bibfnamefont{L.}~\bibnamefont{Zdeborova}},
  \bibinfo{journal}{Proc. Natl. Acad. Sci. USA} \textbf{\bibinfo{volume}{104}},
  \bibinfo{pages}{10318} (\bibinfo{year}{2007}).

\bibitem[{\citenamefont{Selman et~al.}(1996)\citenamefont{Selman, Kautz, and
  Cohen}}]{Selman-Kautz-Cohen-1996}
\bibinfo{author}{\bibfnamefont{B.}~\bibnamefont{Selman}},
  \bibinfo{author}{\bibfnamefont{H.}~\bibnamefont{Kautz}}, \bibnamefont{and}
  \bibinfo{author}{\bibfnamefont{B.}~\bibnamefont{Cohen}}, in
  \emph{\bibinfo{booktitle}{Cliques, Coloring, and Satisfiability}}, edited by
  \bibinfo{editor}{\bibfnamefont{D.~S.} \bibnamefont{Johnson}}
  \bibnamefont{and} \bibinfo{editor}{\bibfnamefont{M.~A.} \bibnamefont{Trick}}
  (\bibinfo{publisher}{Ameri. Math. Society}, \bibinfo{address}{Providence,
  RI}, \bibinfo{year}{1996}), vol.~\bibinfo{volume}{26} of
  \emph{\bibinfo{series}{DIMACS Series in Discrete Mathematics and Theoretical
  Computer Science}}, pp. \bibinfo{pages}{521--532}.

\bibitem[{\citenamefont{Alava et~al.}(2008)\citenamefont{Alava, Ardelius,
  Aurell, Kaski, Krishnamurthy, Orponen, and Seitz}}]{Alava-etal-2008}
\bibinfo{author}{\bibfnamefont{M.}~\bibnamefont{Alava}},
  \bibinfo{author}{\bibfnamefont{J.}~\bibnamefont{Ardelius}},
  \bibinfo{author}{\bibfnamefont{E.}~\bibnamefont{Aurell}},
  \bibinfo{author}{\bibfnamefont{P.}~\bibnamefont{Kaski}},
  \bibinfo{author}{\bibfnamefont{S.}~\bibnamefont{Krishnamurthy}},
  \bibinfo{author}{\bibfnamefont{P.}~\bibnamefont{Orponen}}, \bibnamefont{and}
  \bibinfo{author}{\bibfnamefont{S.}~\bibnamefont{Seitz}},
  \bibinfo{journal}{Proc. Natl. Acad. Sci. USA} \textbf{\bibinfo{volume}{105}},
  \bibinfo{pages}{15253} (\bibinfo{year}{2008}).

\bibitem[{\citenamefont{M{\'{e}}zard
  et~al.}(2005{\natexlab{a}})\citenamefont{M{\'{e}}zard, Palassini, and
  Rivoire}}]{Mezard-etal-2005}
\bibinfo{author}{\bibfnamefont{M.}~\bibnamefont{M{\'{e}}zard}},
  \bibinfo{author}{\bibfnamefont{M.}~\bibnamefont{Palassini}},
  \bibnamefont{and} \bibinfo{author}{\bibfnamefont{O.}~\bibnamefont{Rivoire}},
  \bibinfo{journal}{Phys. Rev. Lett.} \textbf{\bibinfo{volume}{95}},
  \bibinfo{pages}{200202} (\bibinfo{year}{2005}{\natexlab{a}}).

\bibitem[{\citenamefont{M{\'{e}}zard
  et~al.}(2005{\natexlab{b}})\citenamefont{M{\'{e}}zard, Mora, and
  Zecchina}}]{Mezard-etal-2005-a}
\bibinfo{author}{\bibfnamefont{M.}~\bibnamefont{M{\'{e}}zard}},
  \bibinfo{author}{\bibfnamefont{T.}~\bibnamefont{Mora}}, \bibnamefont{and}
  \bibinfo{author}{\bibfnamefont{R.}~\bibnamefont{Zecchina}},
  \bibinfo{journal}{Phys. Rev. Lett.} \textbf{\bibinfo{volume}{94}},
  \bibinfo{pages}{197205} (\bibinfo{year}{2005}{\natexlab{b}}).

\bibitem[{\citenamefont{Seitz et~al.}(2005)\citenamefont{Seitz, Alava, and
  Orponen}}]{Seitz-Alava-Orponen-2005}
\bibinfo{author}{\bibfnamefont{S.}~\bibnamefont{Seitz}},
  \bibinfo{author}{\bibfnamefont{M.}~\bibnamefont{Alava}}, \bibnamefont{and}
  \bibinfo{author}{\bibfnamefont{P.}~\bibnamefont{Orponen}},
  \bibinfo{journal}{J. Stat. Mech.: Theor. Exp.} p. \bibinfo{pages}{P06006}
  (\bibinfo{year}{2005}).

\bibitem[{\citenamefont{Krzakala and Kurchan}(2007)}]{Krzakala-Kurchan-2007}
\bibinfo{author}{\bibfnamefont{F.}~\bibnamefont{Krzakala}} \bibnamefont{and}
  \bibinfo{author}{\bibfnamefont{J.}~\bibnamefont{Kurchan}},
  \bibinfo{journal}{Phys. Rev. E} \textbf{\bibinfo{volume}{76}},
  \bibinfo{pages}{021122} (\bibinfo{year}{2007}).

\bibitem[{\citenamefont{Ardelius and
  Zdeborova}(2008)}]{Ardelius-Zdeborova-2008}
\bibinfo{author}{\bibfnamefont{J.}~\bibnamefont{Ardelius}} \bibnamefont{and}
  \bibinfo{author}{\bibfnamefont{L.}~\bibnamefont{Zdeborova}},
  \bibinfo{journal}{Phys. Rev. E} \textbf{\bibinfo{volume}{78}},
  \bibinfo{pages}{040101(R)} (\bibinfo{year}{2008}).

\bibitem[{\citenamefont{Montanari and
  Semerjian}(2006{\natexlab{a}})}]{Montanari-Semerjian-2006}
\bibinfo{author}{\bibfnamefont{A.}~\bibnamefont{Montanari}} \bibnamefont{and}
  \bibinfo{author}{\bibfnamefont{G.}~\bibnamefont{Semerjian}},
  \bibinfo{journal}{J. Stat. Phys.} \textbf{\bibinfo{volume}{124}},
  \bibinfo{pages}{103} (\bibinfo{year}{2006}{\natexlab{a}}).

\bibitem[{\citenamefont{Montanari and
  Semerjian}(2006{\natexlab{b}})}]{Montanari-Semerjian-2006b}
\bibinfo{author}{\bibfnamefont{A.}~\bibnamefont{Montanari}} \bibnamefont{and}
  \bibinfo{author}{\bibfnamefont{G.}~\bibnamefont{Semerjian}},
  \bibinfo{journal}{J. Stat. Phys.} \textbf{\bibinfo{volume}{125}},
  \bibinfo{pages}{23} (\bibinfo{year}{2006}{\natexlab{b}}).

\bibitem[{\citenamefont{Li et~al.}(2009)\citenamefont{Li, Ma, and
  Zhou}}]{Li-Ma-Zhou-2009}
\bibinfo{author}{\bibfnamefont{K.}~\bibnamefont{Li}},
  \bibinfo{author}{\bibfnamefont{H.}~\bibnamefont{Ma}}, \bibnamefont{and}
  \bibinfo{author}{\bibfnamefont{H.}~\bibnamefont{Zhou}},
  \bibinfo{journal}{Phys. Rev. E} \textbf{\bibinfo{volume}{79}},
  \bibinfo{pages}{031102} (\bibinfo{year}{2009}).

\bibitem[{\citenamefont{Jain and Dubes}(1988)}]{Jain-Dubes-1988}
\bibinfo{author}{\bibfnamefont{A.~K.} \bibnamefont{Jain}} \bibnamefont{and}
  \bibinfo{author}{\bibfnamefont{R.~C.} \bibnamefont{Dubes}},
  \emph{\bibinfo{title}{Algorithms for Clustering Data}}
  (\bibinfo{publisher}{Prentice-Hall}, \bibinfo{address}{Englewood Cliffs, NJ,
  USA}, \bibinfo{year}{1988}).

\bibitem[{\citenamefont{Barthel and Hartmann}(2004)}]{Barthel-Hartmann-2004}
\bibinfo{author}{\bibfnamefont{W.}~\bibnamefont{Barthel}} \bibnamefont{and}
  \bibinfo{author}{\bibfnamefont{A.~K.} \bibnamefont{Hartmann}},
  \bibinfo{journal}{Phys. Rev. E} \textbf{\bibinfo{volume}{70}},
  \bibinfo{pages}{066120} (\bibinfo{year}{2004}).

\bibitem[{\citenamefont{{Dall'Asta} et~al.}(2008)\citenamefont{{Dall'Asta},
  Ramezanpour, and Zecchina}}]{DallAsta-etal-2008}
\bibinfo{author}{\bibfnamefont{L.}~\bibnamefont{{Dall'Asta}}},
  \bibinfo{author}{\bibfnamefont{A.}~\bibnamefont{Ramezanpour}},
  \bibnamefont{and} \bibinfo{author}{\bibfnamefont{R.}~\bibnamefont{Zecchina}},
  \bibinfo{journal}{Phys. Rev. E} \textbf{\bibinfo{volume}{77}},
  \bibinfo{pages}{031118} (\bibinfo{year}{2008}).

\bibitem[{\citenamefont{M{\'{e}}zard and Parisi}(2001)}]{Mezard-Parisi-2001}
\bibinfo{author}{\bibfnamefont{M.}~\bibnamefont{M{\'{e}}zard}}
  \bibnamefont{and} \bibinfo{author}{\bibfnamefont{G.}~\bibnamefont{Parisi}},
  \bibinfo{journal}{Eur. Phys. J. B} \textbf{\bibinfo{volume}{20}},
  \bibinfo{pages}{217} (\bibinfo{year}{2001}).

\bibitem[{\citenamefont{Pearl}(1988)}]{Pearl-1988}
\bibinfo{author}{\bibfnamefont{J.}~\bibnamefont{Pearl}},
  \emph{\bibinfo{title}{Probabilistic Reasoning in Intelligent Systems:
  Networks of Plausible Inference}} (\bibinfo{publisher}{Morgan Kaufmann},
  \bibinfo{address}{San Franciso, CA, USA}, \bibinfo{year}{1988}).

\bibitem[{\citenamefont{Montanari et~al.}(2008)\citenamefont{Montanari,
  {Ricci-Tersenghi}, and Semerjian}}]{Montanari-etal-2008}
\bibinfo{author}{\bibfnamefont{A.}~\bibnamefont{Montanari}},
  \bibinfo{author}{\bibfnamefont{F.}~\bibnamefont{{Ricci-Tersenghi}}},
  \bibnamefont{and}
  \bibinfo{author}{\bibfnamefont{G.}~\bibnamefont{Semerjian}},
  \bibinfo{journal}{J. Stat. Mech.: Theor. Exper.} p. \bibinfo{pages}{P04004}
  (\bibinfo{year}{2008}).

\bibitem[{\citenamefont{Zhou}(2008)}]{Zhou-2008}
\bibinfo{author}{\bibfnamefont{H.}~\bibnamefont{Zhou}}, \bibinfo{journal}{Phys.
  Rev. E} \textbf{\bibinfo{volume}{77}}, \bibinfo{pages}{066102}
  (\bibinfo{year}{2008}).

\bibitem[{\citenamefont{Bouchaud and Dean}(1995)}]{Bouchaud-Dean-1995}
\bibinfo{author}{\bibfnamefont{J.-P.} \bibnamefont{Bouchaud}} \bibnamefont{and}
  \bibinfo{author}{\bibfnamefont{D.~S.} \bibnamefont{Dean}},
  \bibinfo{journal}{J. Phys. I France} \textbf{\bibinfo{volume}{5}},
  \bibinfo{pages}{265} (\bibinfo{year}{1995}).

\bibitem[{not()}]{note3}
\bibinfo{note}{The index S-$i$ (with $i=1, 2, \ldots, 1000$) of a sampled
  solution is equal to the horizontal and vertical position of this solution in
  the plotted Hamming distance matrix. For two solutions S-$i$ and S-$j$ with
  $i< j$, S-$i$ may not necessarily be sampled earlier than S-$j$.}

\bibitem[{\citenamefont{Ciliberti et~al.}(2007)\citenamefont{Ciliberti, Martin,
  and Wagner}}]{Ciliberti-Martin-Wagner-2007-b}
\bibinfo{author}{\bibfnamefont{S.}~\bibnamefont{Ciliberti}},
  \bibinfo{author}{\bibfnamefont{O.~C.} \bibnamefont{Martin}},
  \bibnamefont{and} \bibinfo{author}{\bibfnamefont{A.}~\bibnamefont{Wagner}},
  \bibinfo{journal}{PLoS Comput. Biol.} \textbf{\bibinfo{volume}{3}},
  \bibinfo{pages}{e15} (\bibinfo{year}{2007}).

\end{thebibliography}

\end{document}